\documentclass[aps,prb,twocolumn,showpacs,preprintnumbers,amsmath,amssymb,superscriptaddress]{revtex4-1}

\usepackage{graphicx}
\usepackage{dcolumn}
\usepackage{amsmath,mathrsfs,amssymb,amsthm}
\usepackage{hhline}
\usepackage{wasysym,enumerate,color}
\usepackage{epstopdf}
\usepackage{verbatim}
\usepackage{hyperref}
\usepackage{color}
\usepackage{ulem} 

\renewcommand{\emph}[1]{\textit{#1}} 

\definecolor{darkgreen}{rgb}{0,0.5,0}
\definecolor{purple}{rgb}{0.35,0,0.35}
\definecolor{orange}{rgb}{1,0.5,0}
\definecolor{darkred}{rgb}{.7,0,0}
\definecolor{darkblue}{rgb}{0,0,.3}
\definecolor{grey}{rgb}{.6,.6,.6}
\definecolor{dimgreen}{rgb}{0.2,0.6,0.1}
\hypersetup{colorlinks,linkcolor=blue,urlcolor=blue,citecolor=blue}

\newcommand{\be}{\begin{equation}}
\newcommand{\ee}{\end{equation}}
\newcommand{\bea}{\begin{eqnarray}}
\newcommand{\eea}{\end{eqnarray}}

\newcommand{\s}{{\sigma}}

\newcommand{\bk}{{\mathbf k}}

\newcommand{\tg}{{\tilde \gamma}}

\begin{document}

\title{Finite-frequency thermoelectric response in strongly correlated quantum dots}

\author{Razvan Chirla}
\email{chirlarazvan@yahoo.com}
\affiliation{Department of Physics, University of Oradea, 410087, Oradea, Romania}
\author{C\u at\u alin  Pa\c scu Moca}
\affiliation{Department of Physics, University of Oradea, 410087, Oradea, Romania}
\affiliation{BME-MTA Exotic Quantum Phase Group, Institute of Physics, Budapest University of Technology and Economics,
H-1521 Budapest, Hungary
}
\date{\today}


\begin{abstract}
We investigate the finite-frequency thermal transport through a quantum dot subject to 
strong interactions, 
by providing an exact, nonperturbative formalism that allows us to carry out 
a systematic analysis of the thermopower at any frequency. Special emphasis is put on the dc and high-frequency limits. We demonstrate that, in the Kondo regime, the ac thermopower is characterized by a universal function that we determine numerically.
\end{abstract}


\maketitle


\section{Introduction}\label{sec:Intro}

In the quest to find the most energy-efficient systems and devices, 
thermal generation of currents in nanometer-size structures which can be manipulated by
electric fields, may offer one of the best paths to 
follow.~\cite{Sales.96, Snyder.08, Heremans.08} 
 Quantum dots (QD) are among the best candidates, 
since they are highly tunable. Moreover, they are characterized by an enhanced figure of merit, as a result
of the converging effect of reduced spatial dimensionality, that minimizes 
the phonon thermal conductivity, and an increased electronic density of states. 
So far, they can be used as thermoelectric power generators or 
coolers~\cite{Humphrey.02}, and
when embedded into bulk materials or nanowires,
a structure with large thermopower coefficient, $S$,  
is obtained.~\cite{Harman.02, Wang.08}
The same environment, however, is fundamentally enhancing the interaction between  the 
electrons, generating strong correlations and other dynamical effects. 
By doing detailed measurements in QDs, it has been 
shown that the oscillating behavior of the thermopower as a function of the gate voltage
might carry information on interactions present in the system.\cite{Svensson.12} 
On the theoretical side, the thermoelectric problem in quantum dots is also 
of considerable interest: First, a perturbative calculation 
valid for weakly interacting QDs was  presented  
in Ref.~\onlinecite{Beenakker.92}. 
Later on, in Refs.~\onlinecite{Boese.01, Scheibner.05}
the thermopower of a Kondo correlated dot was computed. 
Quite recently, by using the numerical renormalization group 
approach (NRG) approach, the thermoelectric 
properties of a strongly correlated dot 
modeled in terms of the Anderson model, were investigated systematically.~\cite{Costi.10}
Other more exotic systems such as the SU(4) Kondo state~\cite{Bas.12}
or double-dot systems\cite{Trocha.12} were also studied, 
but, so far, mostly static effects have been addressed~\cite{Donsa.13} and only few theoretical and experimental studies  were focused on dynamical effects.~\cite{Lopez.13} 
\begin{figure}[tbh]
\includegraphics[width=0.9\columnwidth]{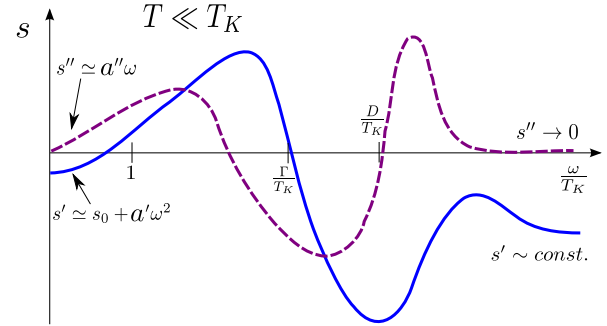}
\caption{(Color online) Sketch of the real and imaginary parts of the 
universal function $s(\omega/T_K, T/T_K)$ in the Kondo regime, 
at a fixed temperature, $T\ll T_K$. The 
coefficients $a'$ and $a''$ depend on temperature as $\sim 1/T^2$.
See also Eq.~\eqref{Eq:un} for the functional dependence of $s$.  
}
\label{fig:s0}
\end{figure}
In contrast, other transport quantities, such as the 
usual differential conductance or the noise, 
have been investigated at various frequencies.~\cite{Sindel.05, Blanter.00, Moca.11}
Consequently, new interesting physics has emerged: It was found that the
modulation of the gate voltages suppresses the Kondo temperature~\cite{Kogan.04} $T_K$, 
and that the frequency-dependent 
emission noise of a quantum dot~\cite{Basset.12} in the Kondo regime, 
at high frequencies, $\hbar\omega \gg k_B T_K$, 
provides information on the system at energy scales which are not accessible by simple 
dc measurements. Furthermore, 
in a slightly different context, i.e., the correlated band models,  
it has been predicted that the thermopower in the high-frequency limit may provide further 
understanding on the thermoelectric transport.~\cite{Shastry.06, Shastry.09, Xu.11}

Motivated by these observations, we consider here the problem of 
thermoelectric response at finite frequencies in a quantum dot 
subject to a strong Coulomb interaction,  and  in particular we shall investigate the dynamical
thermopower $S(\omega)$.  
This quantity characterizes  how a charge current, $I^{(1)}(t)$, is 
generated 
by an infinitesimal time-dependent temperature difference  $\delta T (t)$ across the dot  
and, at the same time, how a
heat current $I^{(2)}(t)$ responds to an infinitesimal voltage drop $\delta V(t)$~\cite{Mahan}
\begin{eqnarray} \label{Eq:S_general}
\left (
\begin{array}{c}
\langle I^{(1)}(t)\rangle \\
\langle I^{(2)}(t)\rangle
\end{array}
\right )
&=& \int  dt' dt''\,T\,S(t-t')\, G (t'-t'')\times
\nonumber \\  
&&\left (
\begin{array}{c}
\delta T(t'')/ T|_{\delta V =0}\\
\delta V(t'')|_{\delta T=0}
\end{array}
\right )\, . 
\end{eqnarray}
As defined, the thermopower itself is not a response function, so it can not be computed directly 
within the linear response theory. Instead, 
the combination $L_{12}(\omega) = T G(\omega)\, S(\omega)$ 
that appears in Eq.~\eqref{Eq:S_general} is a true response function that can be
computed exactly. To get 
the thermopower spectrum $S(\omega)$, aside from $L_{12}$ we also need the 
usual ac conductance, $ G(\omega)=L_{11}(\omega)$. Then, $S(\omega)$ can be
expressed as~\cite{Shastry.09, Luttinger.64}
\begin{equation}\label{Eq:S_omega}
S(\omega)= {1\over T}\left\{ {L_{12}(\omega) \over L_{11}(\omega)}\right \}\, . 
\end{equation}

One of the main results of this work 
is that in the strong coupling (Kondo) regime, i.e., $\max \{\omega, T\}\ll T_K $, the equilibrium ac thermopower takes on a simple, universal 
form, which apart from a phase-dependent prefactor, 
is characterized by a universal function
\begin{equation}
S(\omega, T) \simeq {k_B\over e} \Big ({T\over T_K}\Big )\, s(\omega/T_K, T/T_K)\cot(\delta_0)\,.
\label{Eq:s_intro}
\end{equation}
Here, $\delta_0$ is the phase shift of the electrons 
at  the Fermi level,
and the function $s$ is a complex
universal function that depends exclusively on 
$\omega/T_K$ and $T/T_K$. The prefactor $k_B/e = 8.6 \times 10^{-5} \;V/K$ 
is the unit in which the thermopower is measured. The characteristic features of $s(\omega/T_K, T/T_K)$ 
are sketched in Fig.~\ref{fig:s0}. At a given temperature $T\ll T_K$, and when $\omega \ll T_K$, the real part $s'$ grows 
quadratically with the frequency $s'\simeq s_0 +a'(T)\, \omega^2+\dots$, followed by 
multiple changes of sign at some intermediate frequencies, $\omega_i\sim \{T, \Gamma\}$,  
and becomes constant in the $\omega\to \infty$ limit.
 Its imaginary part
vanishes in the $\omega\to 0$ limit, has a linear dependence $s''\simeq a'' (T)\, \omega+\dots$ 
below the Kondo scale, and vanishes in the $\omega\to \infty$
limit. 

\begin{figure}[h]
\includegraphics[width=0.9\columnwidth]{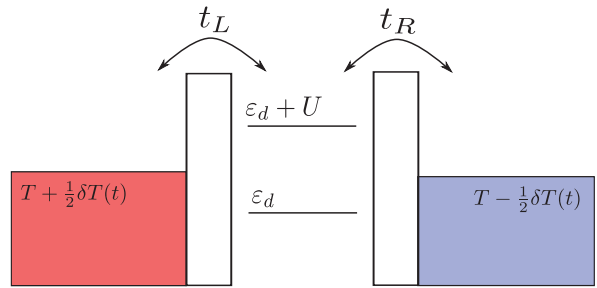}
\caption{(Color online) Sketch of the quantum dot that is coupled 
to two external leads which are assumed to be at different temperatures $T\pm\delta T (t)/2 $.}
\label{fig:dot}
\end{figure}

In  Fig.~\ref{fig:dot}, we present a sketch of the setup. It consists of a quantum dot that is coupled 
to two external leads, $\alpha =\{L,R\}$, that have different temperatures, 
$T_\alpha(t)= T\pm\delta T(t)/2$. The temperature gradient $\delta T(t)$ generates a 
time-dependent current which flows across the dot. Starting from the Kubo formalism and Fourier transforming from time $t$ to 
frequency $\omega$,  we find the ac thermopower, $S(\omega)$. We derive general, exact expressions for $L_{ij}(\omega)$, 
and implicitly for $S(\omega)$, 
which are
valid at any frequency (see Eqs.~\eqref{Eq:S_omega} and  \eqref{Eq:ReLij}). The derivation is then followed by a careful analysis of 
different regimes of interest, 
such as the large-frequency limit, 
$S^*= S(\omega\to \infty)$,  or the conventional low-frequency limit,
$S_0 = S(\omega=0)$~\cite{Costi.10}.
As a technical observation, since we are interested in $S(\omega)$ at any frequency, 
even in the region $\omega\gg D$,
the bandwidth $D$ of the conduction electrons  
must be kept finite in the calculations,
otherwise an unphysical divergence with increasing frequency is present in the spectrum of $L_{12}(\omega)$ (see Sec.~\ref{sec:TP}). 

The paper is organized as follows: In Sec.~\ref{sec:TF}, we introduce the model Hamiltonian
and derive the exact expressions 
for the generalized susceptibilities $L_{ij}(\omega)$ that enter Eq.~\eqref{Eq:S_omega}.
The operators for the charge and heat currents  are discussed in Sec.~\ref{sub:CH}, and 
the results for the ac thermopower are presented in Sec.~\ref{sec:TP}. We 
give the final remarks in Sec.~\ref{sec:C}. Further technical details are discussed in 
Appendices~\ref{app:A} and \ref{app:B}.

%

\section{Theoretical framework}\label{sec:TF}

\subsection{Model Hamiltonian}
In this work, we shall consider the case of a QD described by the Anderson model. It 
consists of a single localized 
orbital that is coupled to two external leads (see Fig.~\ref{fig:dot}). 
The dot can accommodate up to two electrons with strong on-site interaction. 
The Hamiltonian takes the form
\begin{eqnarray}
H & = & \sum_{\sigma}\varepsilon_{d} \;d^{\dagger}_{\sigma}d^{}_{\sigma}+
U\;d^{\dagger}_{\uparrow} d^{}_{\uparrow}d^{\dagger}_{\downarrow} d^{}_{\downarrow}\,\\
&+& \sum_{{\bf k}, \sigma}\sum_{\alpha=L,R}\left (\varepsilon_{\bf k}\;
c^{\dagger}_{\alpha \bf k \sigma}c^{}_{\alpha \bf k \sigma}
+\Big (t_{\alpha{\bf k}}\;
c^{\dagger}_{\alpha \bf k \sigma} d_{\sigma} 
+H.c. \Big )\right )\, , \nonumber
\end{eqnarray}
where $d_{\sigma}$ is the annihilation operator of 
an electron with spin $\sigma$ in the dot, and 
$c^{\dagger}_{\alpha \bf k \sigma}$ is the creation operator of an electron
with momentum $\bf k $ and spin $\sigma$ in lead $\alpha=\{L,R\}$. They satisfy 
the usual anticommutation relations: $\{d_\s,d_{\s'}^{\dagger} \}= \delta_{\s\s'}$ and 
$\{c^{}_{\alpha \bf k \sigma}, c^{\dagger}_{\alpha' \bf k' \sigma'} \} = 
(2\pi)^3\,\delta({\bf k-k'})\,\delta_{\alpha\alpha'}\,\delta_{\sigma\sigma'}$. We treat the leads as having a 
constant density of states $\varrho(\omega)=\varrho_0=1/(2D)$, with $2D$ the bandwidth.  
The tunneling matrix is considered
as being momentum 
independent, $t_{\alpha{\bf k}}= t_{\alpha}$. Its strength is   
characterized  by the usual hybridization function $
\Gamma_{\alpha} = \pi\varrho_0 t_{\alpha}^2$. We define the total hybridization as 
$\Gamma= \sum_{\alpha=\{L,R\}} \Gamma_\alpha$.
The dot itself supports a single orbital with energy 
$\varepsilon_d$ subject to the on-site Coulomb interaction $U$.  Close to the electron-hole 
symmetric configuration, $\varepsilon_d\simeq -U/2$, the dot is in the Kondo regime, characterized
by the Kondo energy scale which is defined as~\cite{Haldane.81}
\begin{equation}
T_K = \sqrt{U\, \Gamma\over 4} e^{\pi 
\varepsilon_d(\varepsilon_d+U)/\Gamma U}. 
\end{equation}

\subsection{Currents and response functions}
\label{sub:CH}
To compute the response functions, we need to define the charge and the heat
currents. We introduce first the charge-  and heat-transfer operators 
 \begin{eqnarray} \label{Eq:QK}
 Q_\alpha = Q^{(1)}_\alpha &=&e\sum_{\bf k, \sigma}c^{\dagger}_{\alpha \bf k \sigma}c^{}_{\alpha \bf k 
\sigma},\nonumber \\
 K_\alpha = Q^{(2)}_\alpha &=&\sum_{\bf k, \sigma}(\varepsilon_{\bf k}-\mu_\alpha)c^{\dagger}_{\alpha \bf k 
\sigma}c^{}_{\alpha \bf k \sigma}\, ,  
\end{eqnarray}
and define the currents as time derivatives of the corresponding charge/heat operators:
\begin{equation}
I_\alpha^{(i)} = (-1)^{i+1}\frac{{\rm d}Q^{(i)}_\alpha}{{\rm d}t}\,. 
\end{equation}
Their explicit expressions 
can be obtained in terms of the equations of motion as:
\begin{eqnarray} \label{Eq:IL1}
I_{\alpha}^{(1)}&= & i\; \frac{e}{\hbar}\; \sum_{\bf k, \sigma}t_{\alpha}\;
c^{\dagger}_{\alpha \bf k \sigma} d_{\sigma} + H.c.\nonumber\\
I_{\alpha}^{(2)} &= &- \frac{i}{\hbar} \sum_{\bf k, \sigma}(\varepsilon_{\bf k}-
\mu_{\alpha})t_{\alpha}\;
c^{\dagger}_{\alpha \bf k \sigma} d_{\sigma} + H.c. \, .
\label{Eq:IL2}
\end{eqnarray}
To avoid a two-channel calculations, it is customary to perform a rotation of the $L/R$ basis to a new one 
 $\{ \alpha_{\bf k \sigma}, \widetilde\alpha_{\bf k \sigma} \}$, 
defined by:
$\alpha_{\bf k \sigma} = \xi_L\,c_{L \bf k \sigma}+\xi_R\,c_{R \bf k \sigma}$
and $\widetilde\alpha_{\bf k \sigma} = \xi_R\,c_{L \bf k \sigma}-\xi_L\,c_{R \bf k 
\sigma}$. This  unitary transformation decouples the 
odd channel,  $\widetilde \alpha_{\bk\s}$, from the dot, so that the dot remains 
coupled only to the even channel, $\alpha_{\bk\s}$. The coefficients are 
$\xi_\alpha=t_\alpha/\sqrt{t_L^2+t_R^2}$ and satisfy  $\xi_L^2+\xi_R^2=1$.
Following this unitary transformation, 
only the interacting part of the Hamiltonian $H_{\rm int}$ changes to
$H_{\rm int}  = t_{\rm eff}\;\sum_{\bf k, \sigma}
\alpha^{\dagger}_{\bf k \sigma} d_{\sigma} + h.c. \, , 
$
with $t_{\rm eff}=\sqrt{t_L^2+t_R^2}$. In what follows we shall consider the 
perfectly symmetric dot, $t_L=t_R=t$. The currents, $I^{(i)}=(I^{(i)}_L-I^{(i)}_R)/2$, 
transform accordingly, and  under equilibrium conditions,
$\mu_{L/R} =0$, in the new basis they are defined as:
\begin{eqnarray} 
I^{(1)}&=&i\,\frac{e\; t_{\rm eff}}{2\sqrt{2}\hbar}\;  \sum_{\bf k, \sigma}\;\widetilde\alpha^{\dagger}_{\bf k \sigma}\, d_{\sigma} + H.c.\label{Eq:I_charge} \\
I^{(2)}&=&-i\, \frac{t_{\rm eff}}{2\sqrt{2}\,\hbar} \sum_{{\bf k}, \sigma}\;\varepsilon_{\bf k}\,\left (\widetilde\alpha^{\dagger}_{\bf k \sigma}\, d_{\sigma} + H.c.\right )\label{Eq:I_heat}\, , 
\end{eqnarray}
and are expressed in terms
of the decoupled channel operators only.  This allows us to obtain exact results 
for the response functions. The currents $I^{(i)}_{L/R}$ also have a symmetrical component, 
which gets subtracted out in the definition of $I^{(i)}$. Within the Kubo formalism,  
the generalized response functions $L_{ij}$ are given by\cite{hewson}
\begin{equation}
L_{ij}(t,t')= -\frac{i}{\hbar}\Theta(t-t')\langle [I^{(i)}(t), Q^{(j)}(t')]\rangle\, ,
\end{equation}
where
\begin{eqnarray}
Q^{(i)}=\frac{Q^{(i)}_L-Q^{(i)}_R}{2}\;.
\end{eqnarray}
We want to express $L_{ij}$ in terms of locally defined operators only, and 
for that we eliminate the charge operators. In Fourier space we obtain
\begin{equation} \label{Eq:L}
L_{ij}(\omega)=-{T_{ij}(\omega)-T_{ij}(0)\over i\, \omega}\, ,
\end{equation}
with $T_{ij}(\omega)$ the Fourier transform of the generalized susceptibility:
\begin{equation} \label{Eq:T}
T_{ij}(t,t')=  -\frac{i}{\hbar}\, \Theta(t-t')\langle[  I^{(i)}(t),I^{(j)}(t') ]\rangle.
\end{equation}
Somewhat similar to the calculation of the ac conductance~\cite{Sindel.05}, 
the calculation of thermopower, $S(\omega)= S'(\omega)+S''(\omega)$, 
reduces to the calculation of 
$A_{\rm d}(\omega)= -\rm Im\, G_{\rm d}^{\rm R}(\omega)/\pi$ - the spectral 
representation of the d-level operators in the dot (see Appendix~\ref{app:B}). 
In the present work $A_{\rm d}(\omega)$ shall be
computed by using the Wilson numerical renormalization 
group approach\cite{Wilson.80.1, Wilson.80.2, Bulla.08} (NRG) as implemented in the 
Flexible-NRG code.~\cite{BudapestNRG} Throughout the NRG calculation, the  
Wilson ratio was fixed to $\Lambda=2$, and we have kept on average 4000 multiplets at each iteration.

\section{Ac thermopower}\label{sec:TP}
\begin{figure}[th]
\includegraphics[width=0.47\columnwidth]{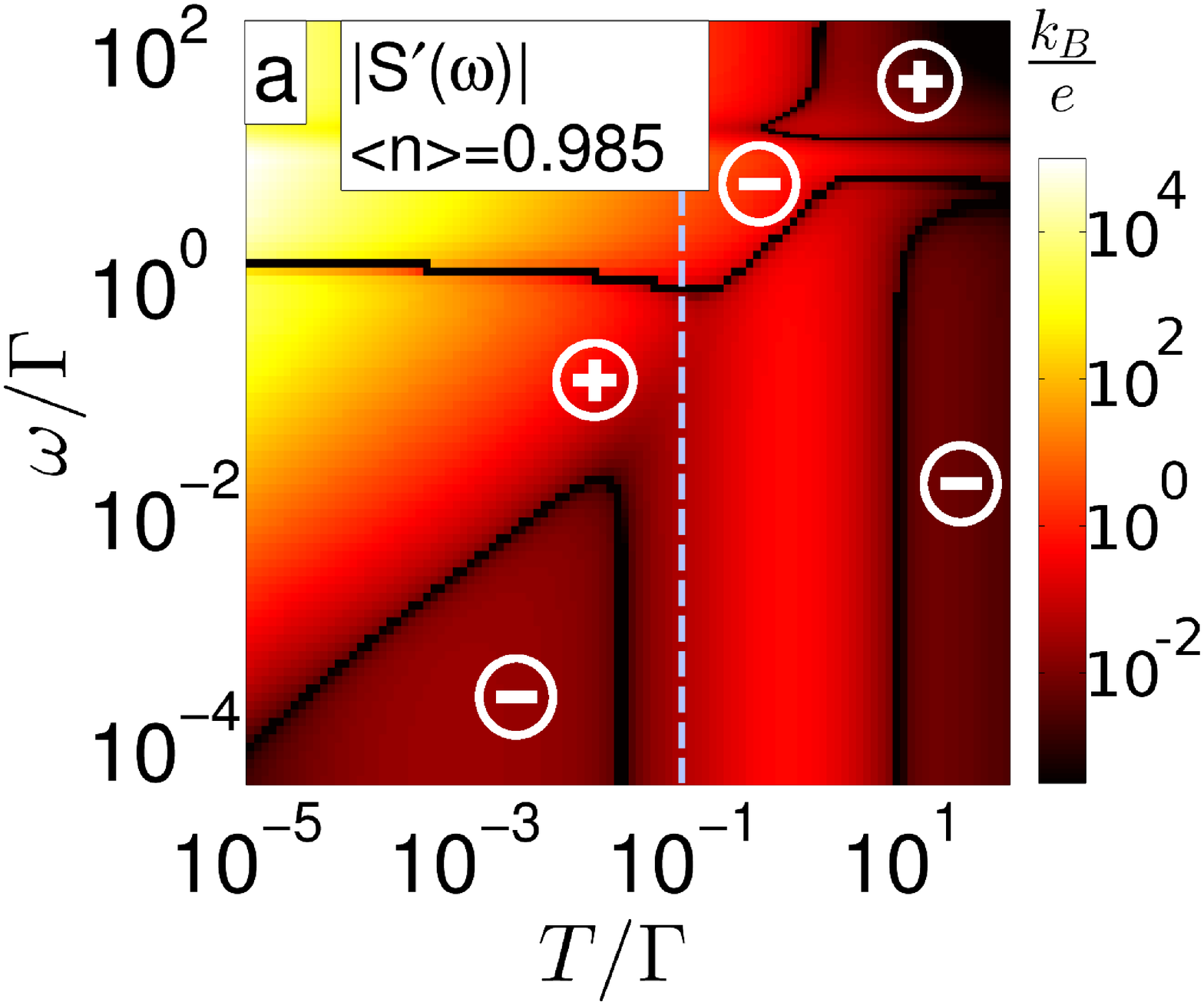}
\includegraphics[width=0.47\columnwidth]{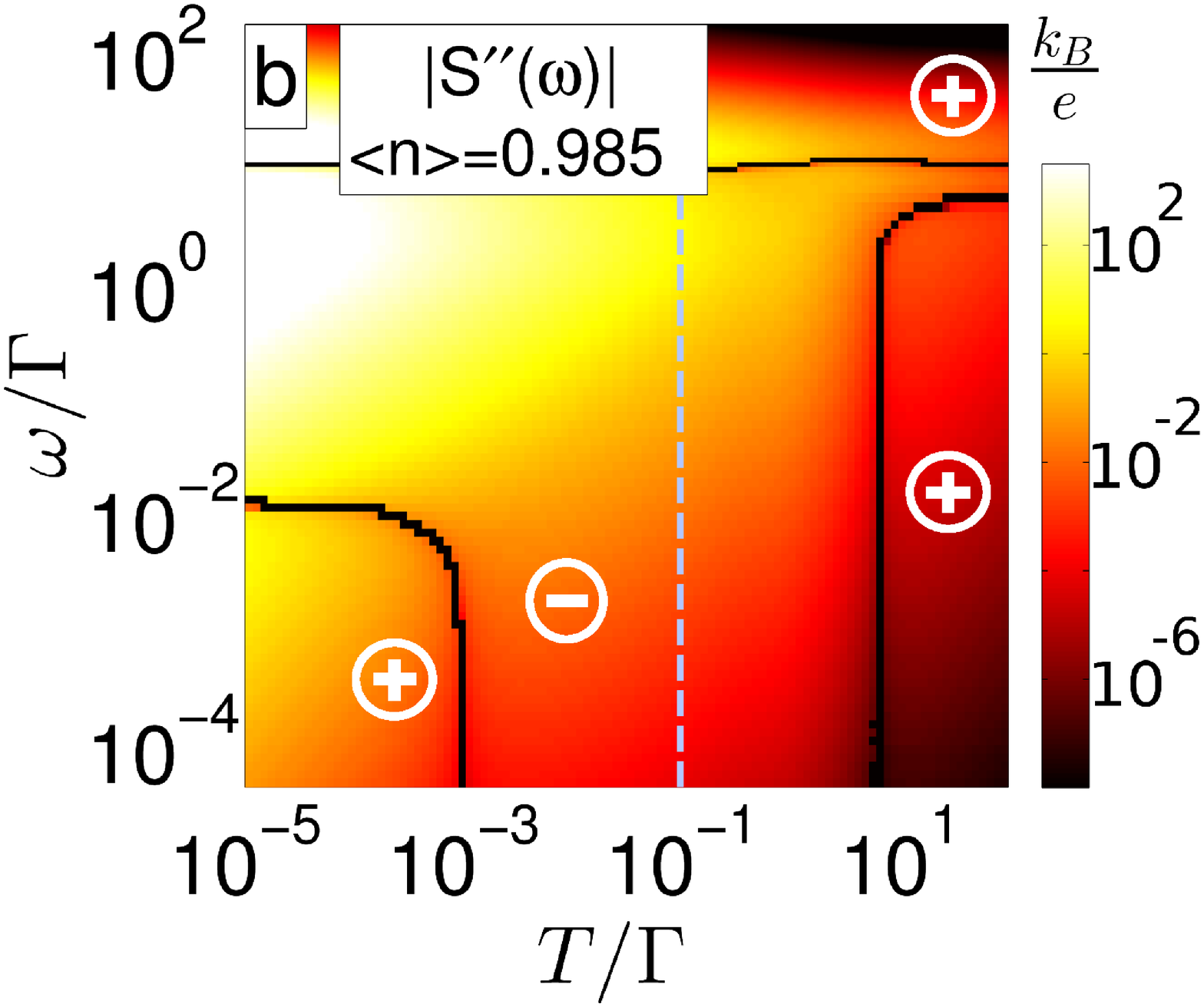}
\includegraphics[width=0.47\columnwidth]{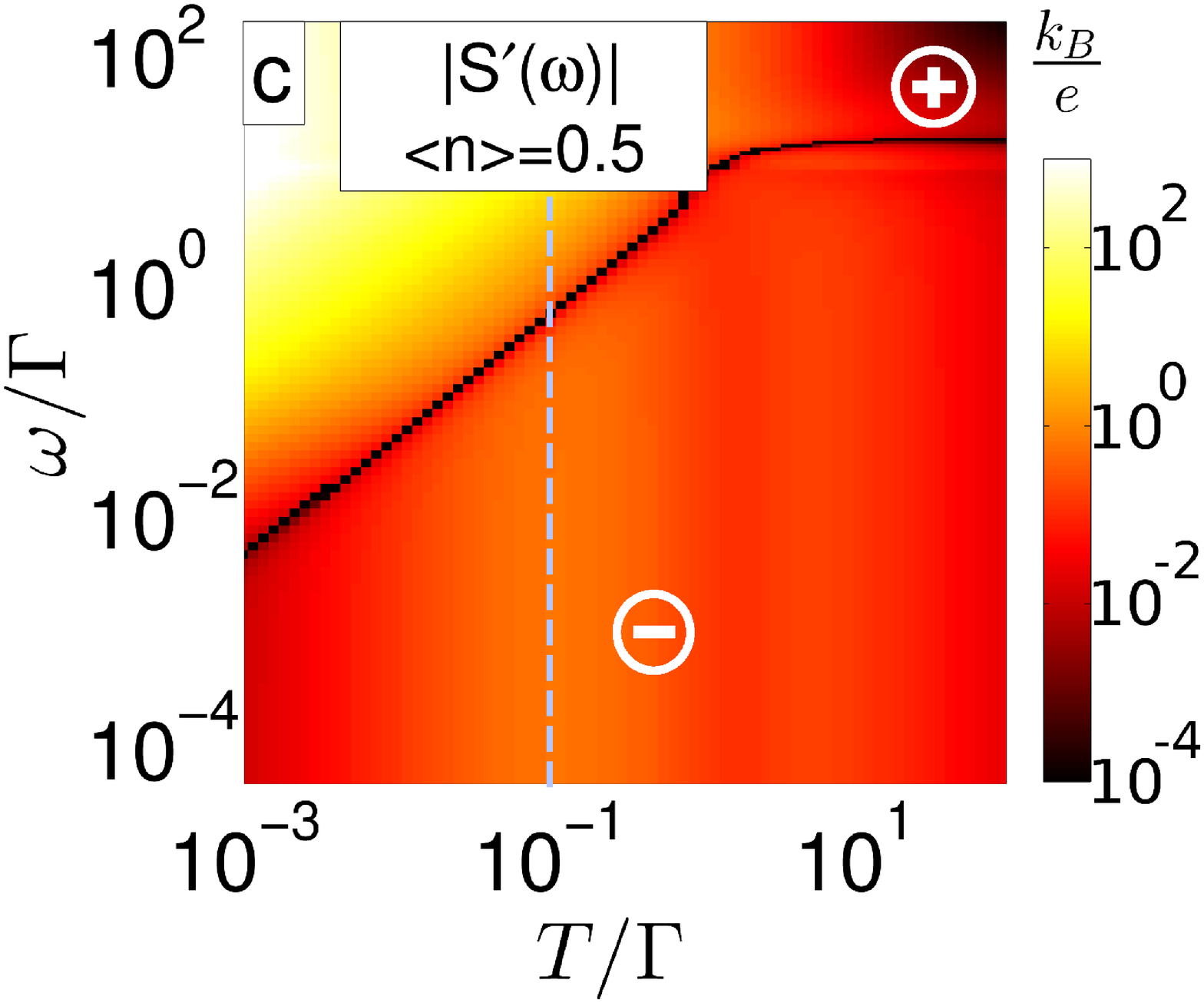}
\includegraphics[width=0.47\columnwidth]{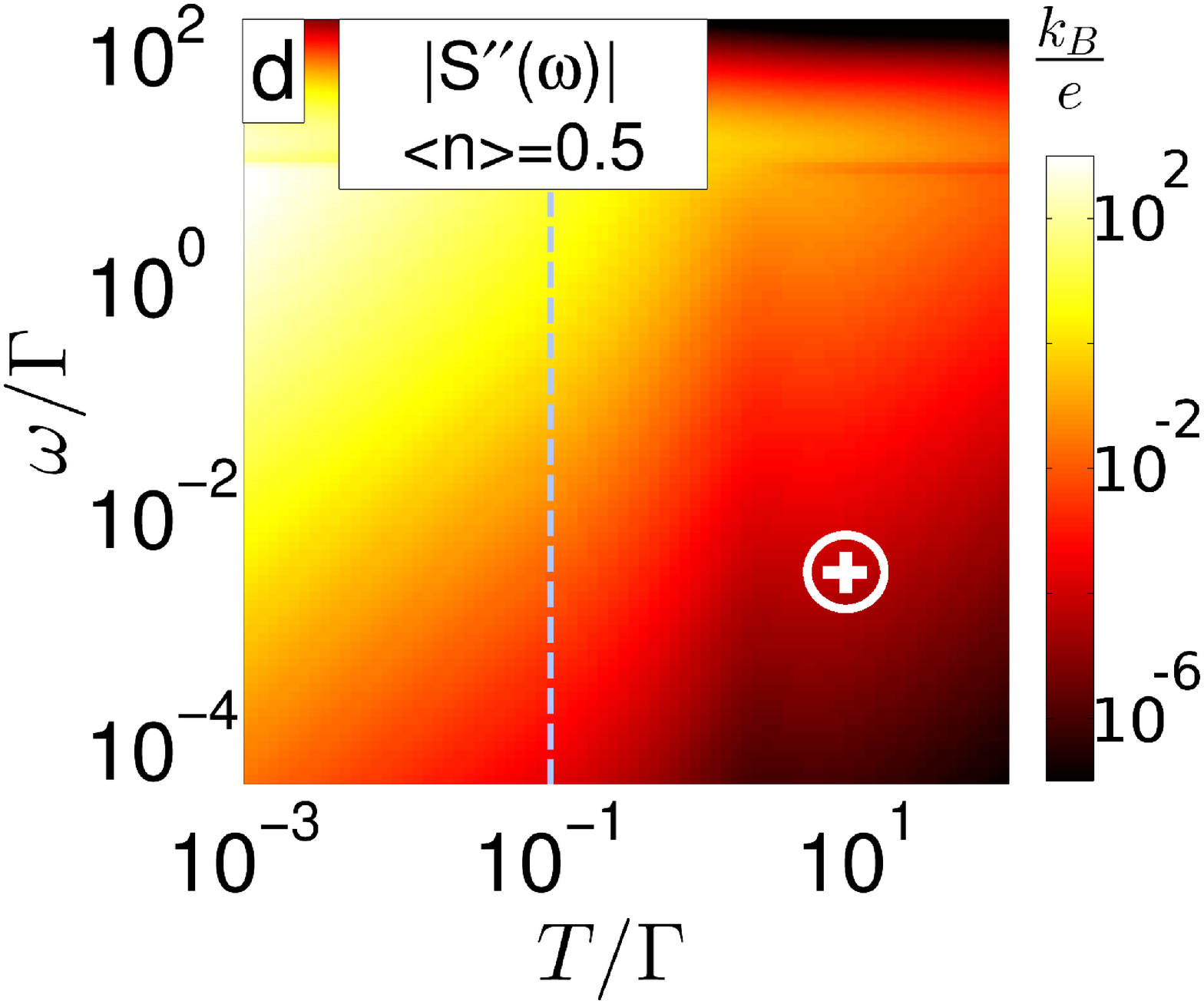}
\includegraphics[width=0.47\columnwidth]{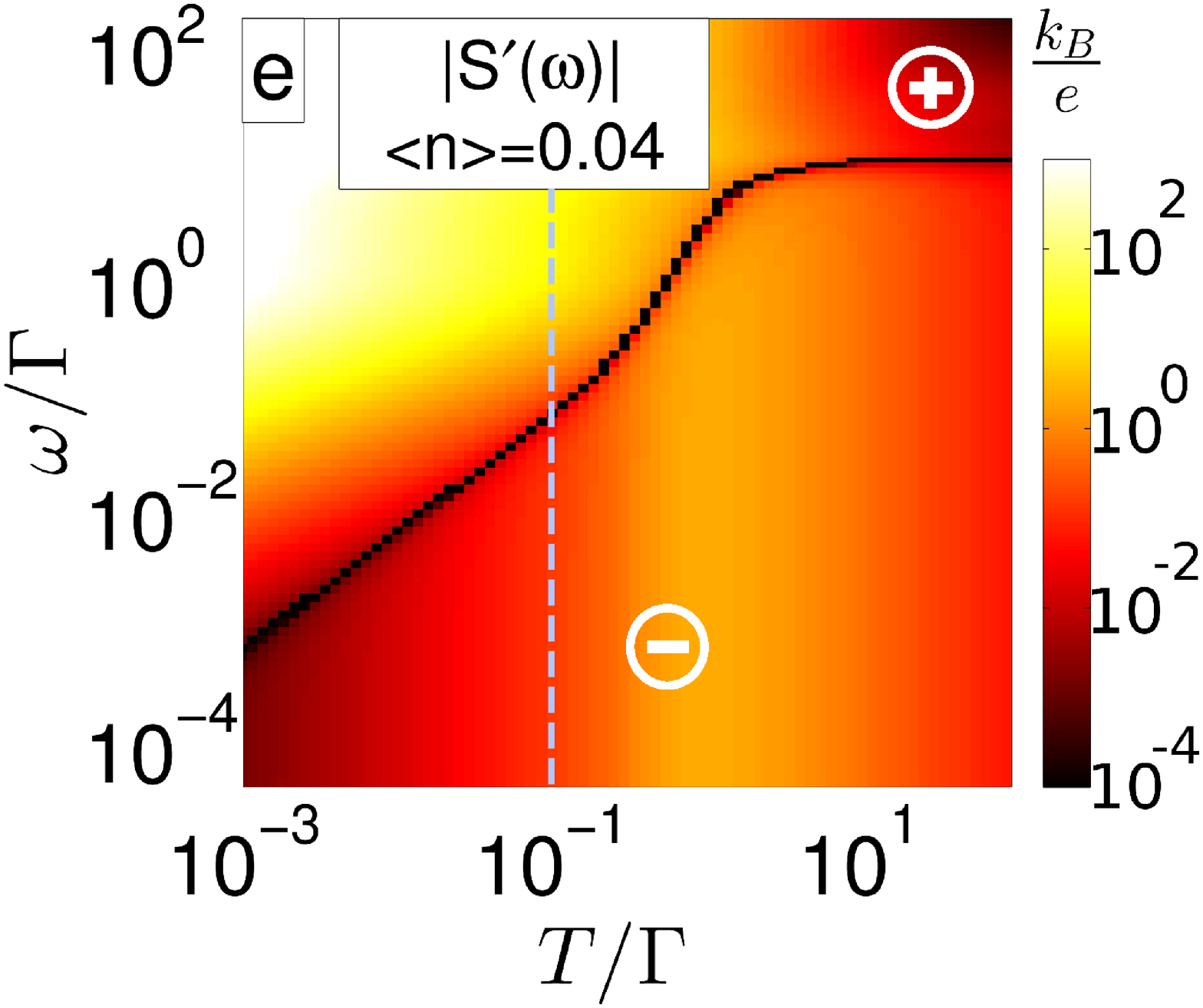}
\includegraphics[width=0.47\columnwidth]{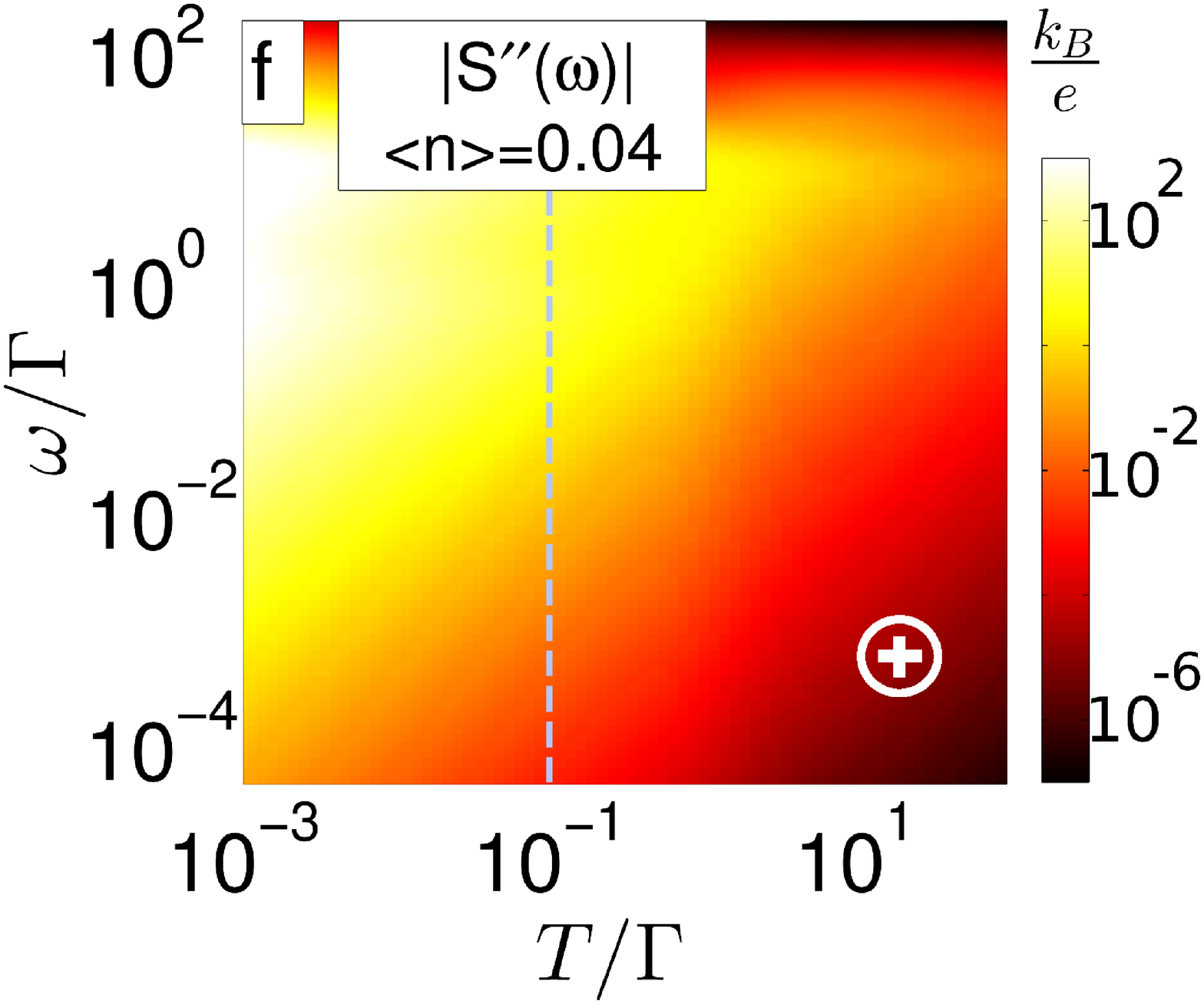}
\includegraphics[clip,width=0.47\columnwidth]{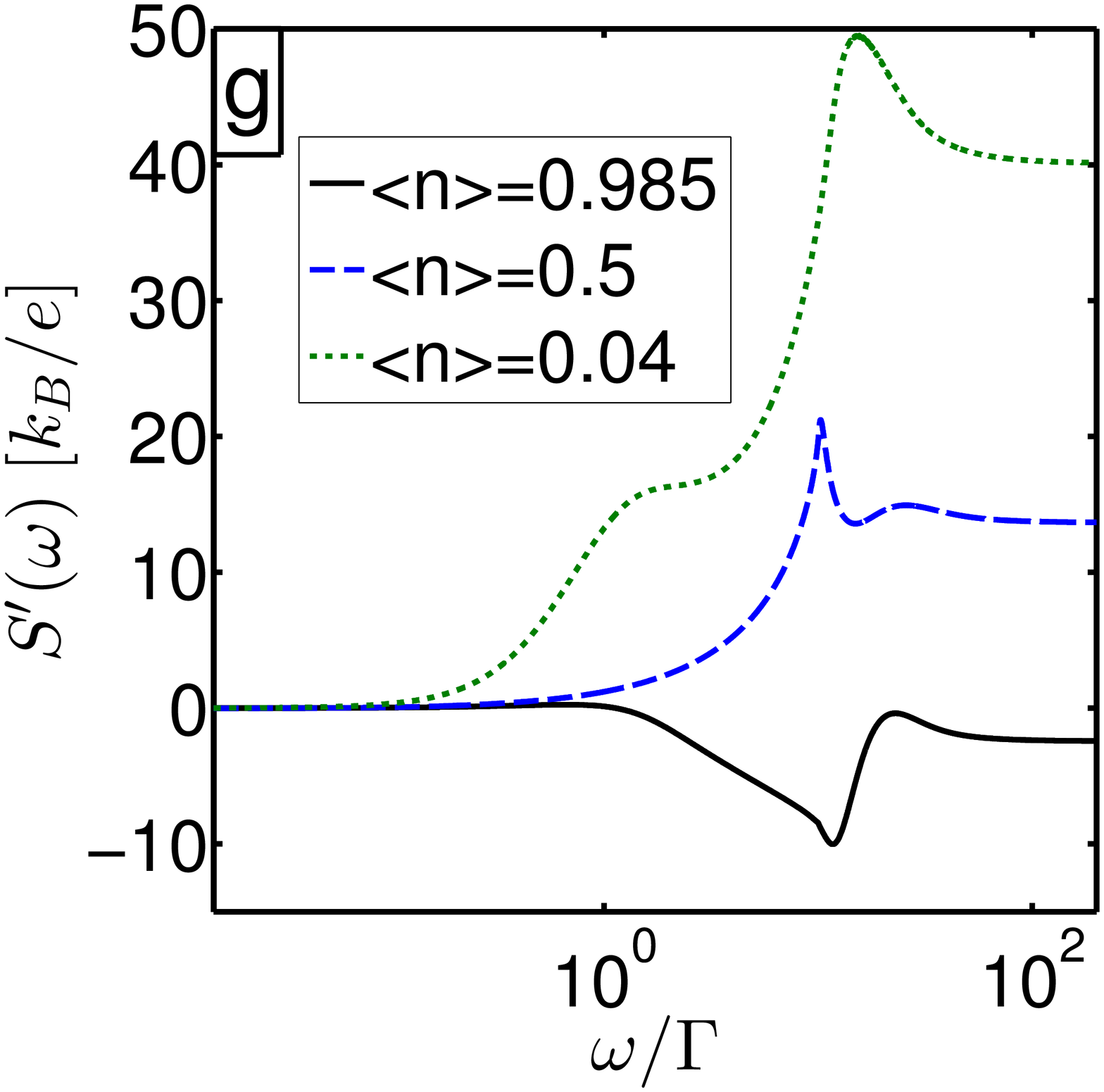}
\includegraphics[clip,width=0.47\columnwidth]{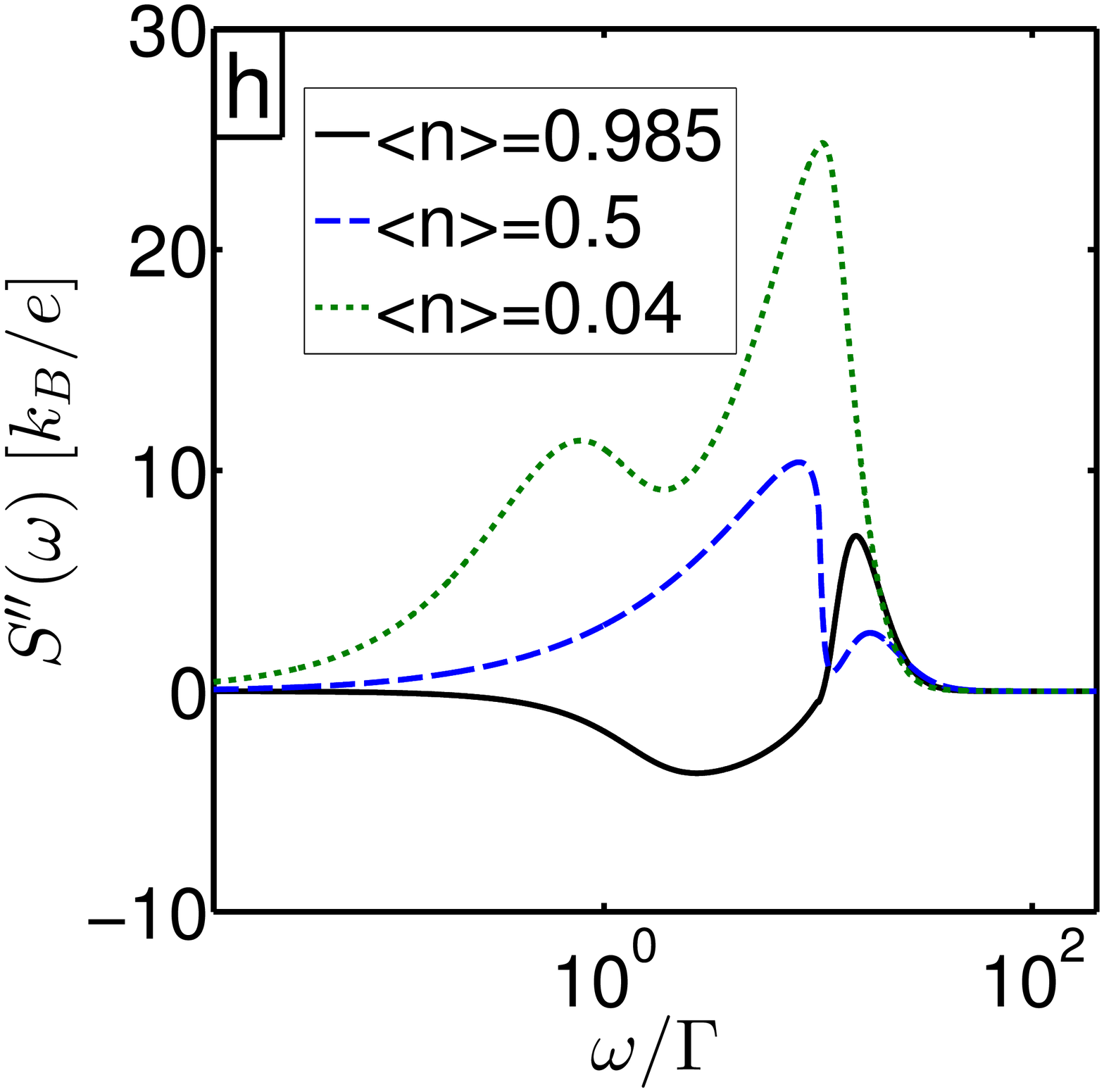}
\caption{(Color online) Density plots for the real and imaginary parts of the thermopower
in the $(T, \omega)$ plane: (a), (b) Kondo regime, (c), (d) mixed valence regime, and (e), (f) 
empty orbital regime. Panels (g) and (h) represent the thermopower  at $T= 0.1\,\Gamma$ for positive 
$\omega$ along the dashed lines in the density plots. The darker lines are the locations of the zeros, and mark 
the positions where the thermopower changes sign. The $(\pm)$ symbols indicate the 
signs of  $S(\omega)$. The NRG parameters used are  $U/\Gamma=10$, $\varepsilon_d/\Gamma = -4\, (n=0.985)$, 
$\varepsilon_d/\Gamma=-0.2\, (n=0.5)$ and $\varepsilon_d/\Gamma = 5\, (n=0.04)$.} 
\label{fig:S}
\end{figure}

Let us now focus on the calculations of the response functions $L_{ij}$ and the 
ac thermopower $S(\omega)$.
In general, $L_{ij}(\omega)$ are
complex functions of $\omega$, as their imaginary parts capture retardation 
effects due to the external excitation. 
The full $\omega$ dependence of the ${\rm Re}\, L_{ij}(\omega)$    
acquires a relatively compact expression in terms of the spectral representation
of the d-level:
\begin{eqnarray} \label{Eq:ReLij}
\textrm{Re}\,L_{ij}(\omega)=\frac{t_{\rm eff}^2}{2\omega \hbar} \left( -\frac{e}{\hbar}\right)^{4-i-j} 
\varrho_0
\int \rm d\omega'\left \{\textrm{Im}G_d^R(\omega') \right. \nonumber \\
\left[ 
(\omega'-\omega)^{i+j-2}\Theta_{\omega'-\omega}f(\omega'-\omega)+\right. \nonumber \\
+ (\omega'+\omega)^{i+j-2}\Theta_{\omega'+\omega}f(\omega')-\nonumber \\
-(\omega'-\omega)^{i+j-2}\Theta_{\omega'-\omega}f(\omega')\nonumber \\
\left. \left.-(\omega'+\omega)^{i+j-2}
\Theta_{\omega'+\omega}f(\omega'+\omega)\right]  \right\}\, , \nonumber\\
\end{eqnarray}
with $\Theta_\omega=\Theta(D-|\omega|)$ and $f(\omega)$ the Fermi-Dirac distribution.  
To get the ac conductance~\cite{Sindel.05}, 
the high energy cut-off $D$ can be safely taken to infinity, as
$L_{11}$ remains a regular function. 
This is not the case for $L_{12}$ 
which diverges at large frequencies when $D\to \infty$, so it is compulsory to have a finite bandwidth
for the conduction electrons.  To get the imaginary part of $L_{ij}$,
the use of the Hilbert transform is unavoidable. We first compute ${\rm Re}\, L_{ij}$, 
and then ${\rm Im}\, L_{ij}$ is obtained by a Kramers-Kr\" onig(KK) transformation
\begin{equation} \label{Eq:KK}
\textrm{Im}\, L_{ij}(\omega)=-\frac{1}{\pi}\int_{-\infty}^{+\infty}\frac{\textrm{Re}
\,L_{ij}(\omega')}{\omega'-\omega}\rm d \omega'
\end{equation}
Although Eq.~\eqref{Eq:ReLij} looks cumbersome, we can interpret it in terms of
inelastic tunneling processes and see these correlation functions as rates 
by which the system absorbs or emits photons~\cite{Basset.12} at frequencies $\omega$. Further details on how to compute $L_{ij}$ are presented 
in Appendix \ref{app:B}. 
With $L_{ij}$ at hand, $S(\omega)$ can be obtained by using Eq.~\eqref{Eq:S_omega}.
By symmetry,  $S'(\omega)= S'(-\omega)$,  is an even function of frequency,
while $S''(\omega)= -S''(-\omega)$, is an odd function. 
The full $\omega$ and $T$ dependence for $S(\omega)$ is displayed in Figs. ~\ref{fig:S} (a)-~\ref{fig:S}(f),  while 
in Figs. ~\ref{fig:S}(g) and ~\ref{fig:S}(h) a cut at constant temperature $T= 0.1\, \Gamma$ is presented.
%
%
\begin{figure}[tbh]
\includegraphics[clip,width=0.9\columnwidth]{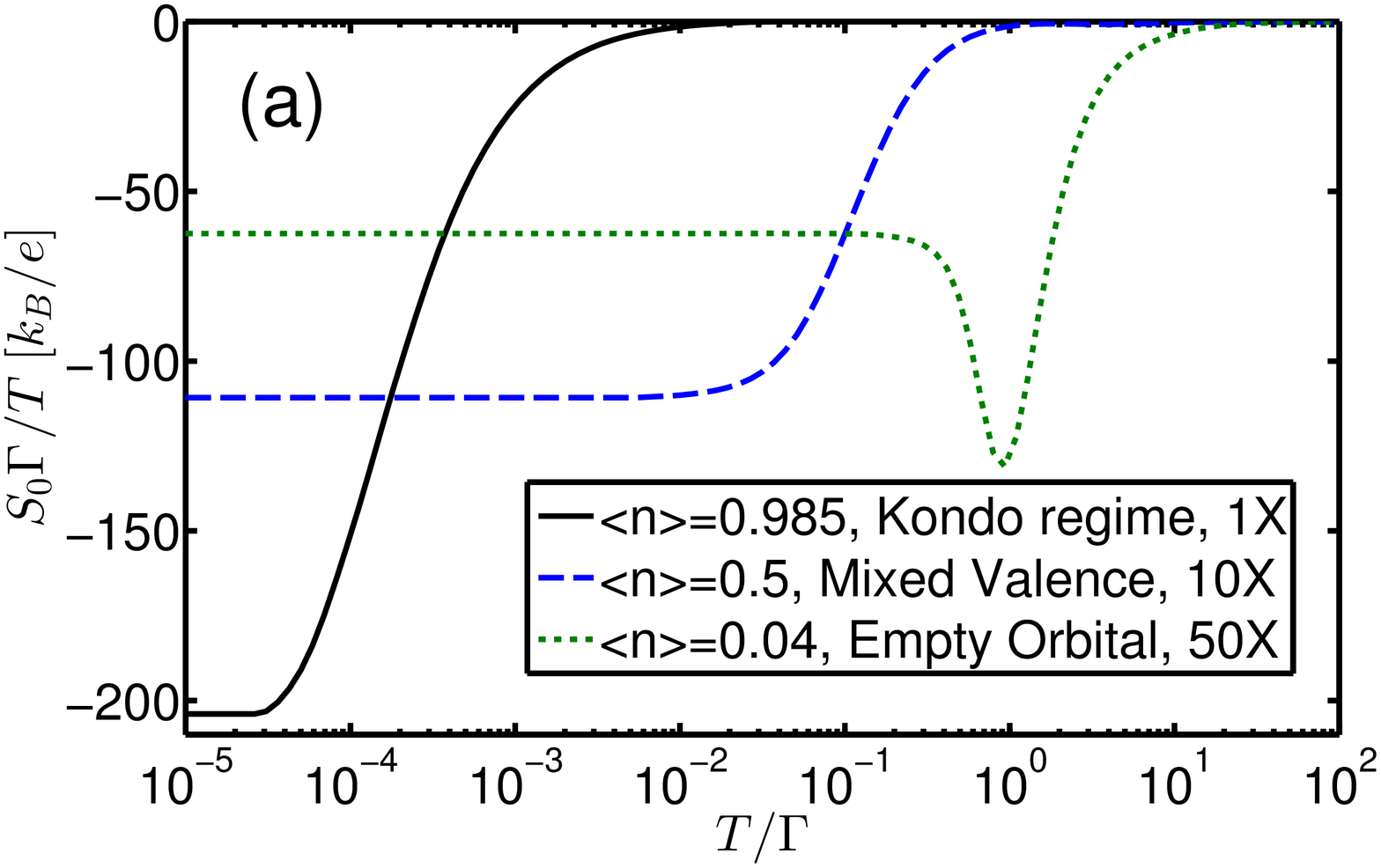}
\includegraphics[clip,width=0.9\columnwidth]{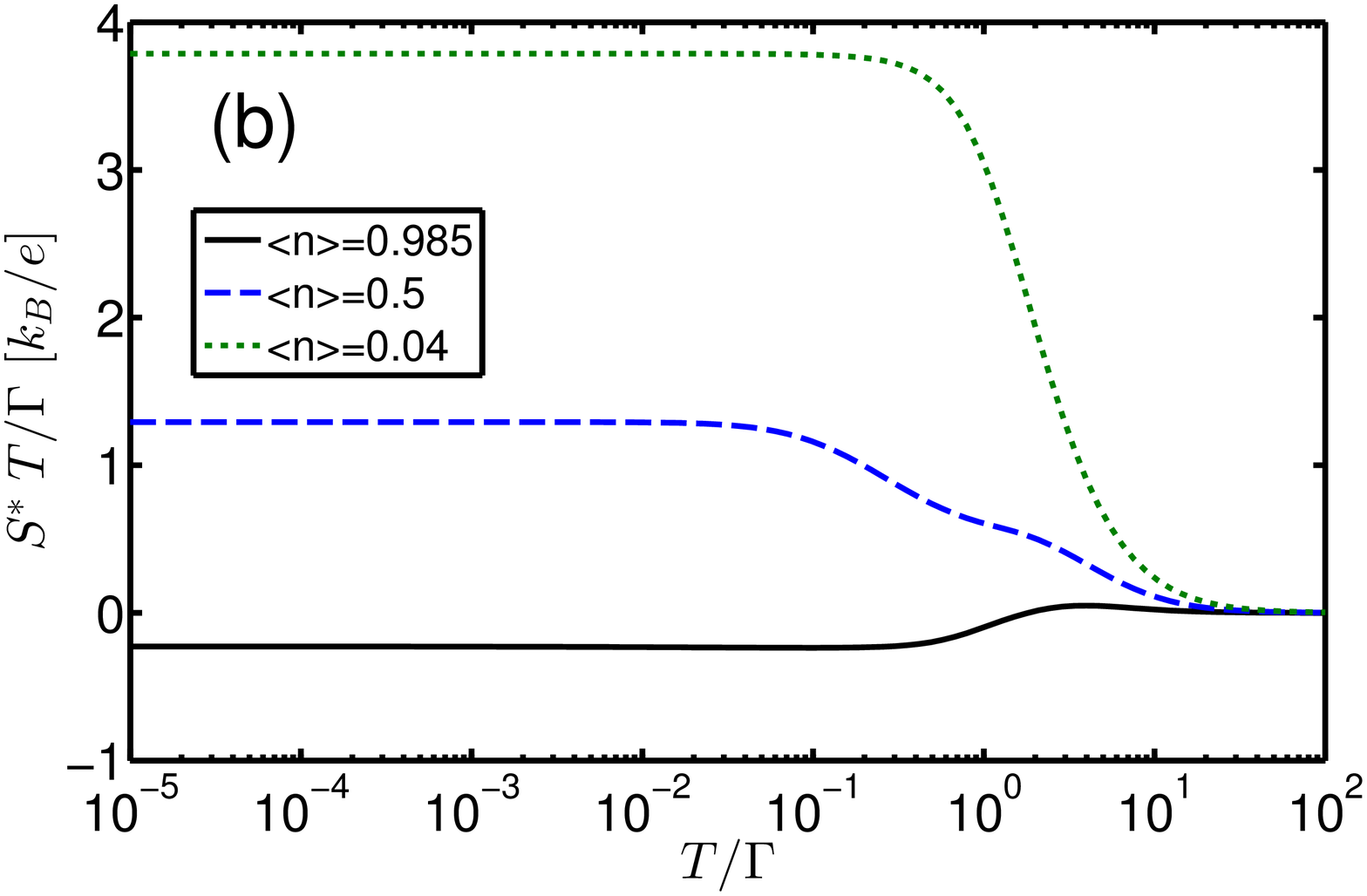}
\caption{(Color online) (a) The dc-thermopower $S_0=S(\omega=0)$
as function of temperature in different regimes. 
(b) Temperature dependence of $S^*=S(\omega \gg D)$.}
\label{fig:S_zero}
\end{figure}
In Figs. ~\ref{fig:S}(a)-~\ref{fig:S}(b), we represent the results for $S(\omega, T)$ when the system is 
in the Kondo regime, where $\epsilon_d\simeq -U/2$, and $\langle n\rangle \simeq 1$. 
Figs. ~\ref{fig:S}(c) and ~\ref{fig:S}(d) present results for $S(\omega, T)$ in 
the mixed valence regime with $\langle n\rangle = 0.5$, while Figs. ~\ref{fig:S}(e) and ~\ref{fig:S}(f) show $S(\omega, T)$ in the empty orbital limit, $\langle n \rangle\ll 1$. 
 By symmetry, when $1<n<2$, the thermopower has the same magnitude, but opposite sign, 
as the role of particles and holes is inverted. At the electron-hole symmetric point, 
i.e., $n=1$, the 
particles and holes move together in the same direction under the temperature gradient,
and the thermopower vanishes exactly.

At very small frequencies and temperatures, $\{\omega, T\} \to 0 $, 
only the quasiparticles very close to the Fermi surface 
give a contribution to the currents flowing through the device, 
so this limit can be understood in terms of the Fermi-liquid picture.~\cite{Nozieres.74}
The Fermi-liquid scale $\Omega_F$, is controlled by either $T_K$ within the Kondo regime, 
or by  $\Gamma$ itself otherwise.  
When $\{\omega, T\}\ll \Omega_F$, the
frequency dependence of $S(\omega)$ is captured by a simple analytical expression
\begin{equation}
S(\omega, T) = S_0(T)+ b'(T) \, {\omega}^2+i\,b''(T)\, {\omega}+\dots\, . 
\end{equation}
Here $S_0(T)<0$, is the dc-thermopower, while $b'$ and $b''$ are 
some coefficients that depend  on temperature. In the Kondo regime, 
$\{b'(T), b''(T)\}\sim 1/T$.
 In our convention, 
positive (negative) $S_0$ corresponds to the situation when charge and 
heat currents flow in the same (opposite) directions. At some intermediate frequencies, 
$\omega\sim T$, $S'(\omega)$ changes sign and becomes positive. In the Kondo regime,
there is another change of sign at a larger, almost constant frequency $\omega_2\simeq \Gamma$, 
and $S'$ becomes negative in the $\omega\gg D$ limit. 
The sign of $S_0(T)$ can be associated with the type of dominant carriers in the system
at that particular energy: hole like carriers correspond to $S_0>0$, and particle like carriers to $S_0<0$. 
The first sign change in $S_0(T)$ can be
understood as follows:  At $T \simeq 0$, the Kondo peak is weighted towards
positive energies (for $n<1$), but as temperature increases, 
it is pulled towards the
negative energy region. Thus, the quantum dot shifts from having 
predominantly particle carriers, to having 
predominantly hole carriers in the window $\sim T$ that contributes to the transport. Whenever the average entropy carried inside this window becomes zero, the thermopower vanishes. 
At a finite frequency $\omega$, inelastic tunneling processes in a window $\sim 2 \omega$ around the Fermi level give additional contributions to the transport, see Eq. (\ref{Eq:ReLij}). The picture gets more complicated by the existence of retardation effects, which lead to finite imaginary parts in the response functions, and consequently affect the zeros of the thermopower.
The high-frequency features in Fig.~\ref{fig:S} at energies  $\omega \sim \{U, D\}$ 
can be associated with Hubbard charging, and eventually band-edge effects. 
In Fig.~\ref{fig:S_zero}, we display the temperature dependence of  $S_0(T)$ and $S^*(T)$.
It has been already shown~\cite{Costi.93} that  in the Kondo regime, when $T\ll T_K$,  
$S_0$ depends linearly on $T$, $S_0\propto  T$. 
This observation carries over to the mixed valence and
empty orbital regimes too, as long as $T\ll \Gamma$. In the opposite limit, 
$T\gg T_K$, $S_0$ shows a change 
of sign at some large temperature, and then decays towards zero.

 In the large-frequency limit $\omega\gg D$, 
 $S^*$ can be evaluated simply as
\begin{equation}\label{Eq:S_star}
S^*(T) = \frac{1}{T}\frac{L_{12}^*}{L_{11}^*}.
\end{equation}
with $L_{ij}^*$ some temperature-dependent coefficients, discussed in Appendix~\ref{app:B} 
(see Eq.~\eqref{Eq:L_star}). 
As $\omega\gg D$,  all states are  involved in transport, so that the only energy scale that 
survives is the bandwidth itself (as long as $\{T_K, U\}\ll D$). Therefore, we expect the features of $S^*(T)$ to carry 
information only on $D$ itself.
In the small-temperature limit, $T\to 0$, the $L_{ij}^*$'s become constants, so 
a divergent behavior $S^* \propto 1/T$ emerges in this limit. 
This is clearly visible in Fig.~\ref{fig:S_zero} (b), where  $S^* T$ becomes constant when $T\ll \Gamma$. 
At intermediate temperatures, $\Gamma<T<D$,  $S^*$ decreases significantly,
and vanishes in the in large $T$ limit.

In what follows we shall focus on the strongly correlated regime.   
When $\{\omega, T\}\leq T_K$, a clear
universal behavior emerges as $S(\omega)$ depends exclusively on 
the $T/T_K$ and $\omega/T_K$ ratios.  
\begin{figure}[tbh]
\includegraphics[clip,width=0.99\columnwidth]{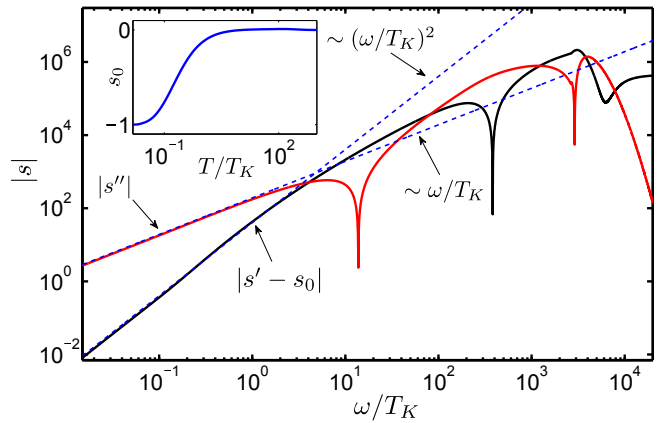}
\caption{(Color online) Universal scaling functions for a filling $\left <n \right>=0.95$ for a 
fixed temperature, $T=0.02\;T_K$. In the inset, we represent the $s_0$ universal function. }
\label{fig:S_scaling}
\end{figure}
It was found previously\cite{Costi.10} that as long as $T\ll T_K$, %
\begin{eqnarray}\label{Eq:s0}
s_0 (T/T_K) &=& \Big (\frac{e}{\pi\gamma\, T}\Big )S_0(T)\,\tan \delta_0\nonumber\\
& = &  \frac{e}{k_B}\Big (\frac{T_K}{T} \Big )S_0(T)\,\left (\frac{\tan \delta_0}{\tg} \right )
\end{eqnarray}
is a universal function that scales with $T/T_K$  up to a filling-dependent phase factor. 
Here, $\delta_0$ is the phase shift at the Fermi level, 
$\delta_0 = \pi\langle n\rangle/2$, with $\langle n\rangle $ the average occupation of the dot,
$\gamma$ is the specific-heat coefficient of the quantum dot which is filling dependent, and 
$\tg = \pi \gamma\, T_K/k_B$ is a dimensionless quantity of the order 1. In the inset of Fig.~\ref{fig:S_scaling} we represent $s_0(T/T_K)$ as a function of $T/T_K$
for a filling $\langle n \rangle =0.95$.
 We extend this analysis  to
finite frequencies where a similar behavior emerges, as
the ac thermopower can be expressed as:
\begin{equation}\label{Eq:s}
S(\omega, T) = {k_B\over e} \Big ({T\over T_K}\Big )\, s(\omega/T_K, T/T_K)\, \tg\,\cot(\delta_0)\,.
\end{equation}
The universal scaling function $s$  depends on $\omega/T_K$ and $T/T_K$ only. 
A sketch with the frequency dependence is presented in Fig.~\ref{fig:s0}, while in  
Fig.~\ref{fig:S_scaling} we represent 
the exact numerical calculation for the frequency dependence of $|s|$. 
In the Fermi-liquid regime, $\{\omega, T\}\ll T_K$, simple analytical expressions
can approximate the real and imaginary parts of $s$:
\begin{eqnarray}\label{Eq:un}
s'\Big({\omega\over T_K}, {T\over T_K}\Big)& =&  s_0 \Big({T\over T_K}\Big) + 
\alpha' \Big({\omega\over T_K}\Big)^2\Big({T_K\over T}\Big)^2+\dots, \nonumber\\
s''\Big({\omega\over T_K}, {T\over T_K}\Big) &= & \alpha'' \Big({\omega\over T_K}\Big)
\Big({T_K\over T}\Big)^2+\dots.
\end{eqnarray}
with $\alpha'$ and $\alpha''$ some coefficients $\sim 1$. 
The $s'\sim \omega^2$ frequency dependence can be related to the  
virtual Kondo transitions from the singlet ground state to the excited states.~\cite{Glazman.03} 
These transitions give for the imaginary part of the T-matrix:
$\textrm{Im}\,{\rm T}(\omega,T)\propto1- {(3\, \omega^2+\pi^2 T^2)}/{T_K^2}$, 
  when $\{\omega, T\}\ll T_K$, 
which in turn introduces corrections of the order  $\sim \omega^2$ and $\sim T^2$ in the 
response functions $\textrm{Re}\,L_{ij}$. Simple analytics then show that 
$s'\sim\omega^2/T^2$. Then, by Hilbert transform, $s''$ is linear in frequency.
This scaling for $s$ extends up to frequencies of the order of $\omega \sim T_K$, 
followed by a 
sign change at some particular frequencies $\omega_i$. 
At very large frequencies $\omega\gg T_K$, $s'$ becomes a constant, while 
$s''\to 0$.


\section{Concluding Remarks}\label{sec:C}

We have studied the finite-frequency thermopower of a quantum dot 
described by the Anderson model. For that we have first 
constructed a general framework 
which allowed us to 
investigate in a non-perturbative manner 
the ac thermopower.  When calculating  the 
ac conductance~\cite{Sindel.05},
 it is safe to take the bandwidth $D\to \infty$, but 
when we address the problem of the ac thermopower, it is compulsory 
to keep $D$ finite.  Although $S(\omega)$ presents 
a relatively rich structure that includes several sign changes, 
in the Fermi liquid regime a simple 
analytical expression is able to capture its behavior over a broad range of
temperatures and frequencies. In the  Kondo regime, the ac thermopower is
characterized by a universal function that 
we have determined numerically. We have also found that the $S_0$ 
and $S^*$ have a markedly different behavior in the low-temperature limit. 

\section*{Acknowledgments}
We would like to thank I. Weymann for carefully reading
our manuscript. This research has been financially supported by UEFISCDI under 
French-Romanian Grant DYMESYS (PN-II-ID-JRP-2011-1 and ANR 2011-IS04-001-01)
and by Hungarian Research Funds under grant Nos. 
K105149, CNK80991, TAMOP-4.2.1/B-09/1/KMR-2010-0002.

\appendix
\section{Green's Functions}\label{app:A}
%
Within the L/R basis transformation, one channel becomes decoupled, and can be treated 
as a non-interacting one. It is simply described by the non-interacting
Hamiltonian $H_{0}  = \sum_{\bf k, \sigma}\varepsilon_{\bf k}\;
\widetilde{\alpha}^{\dagger}_{\bf k \sigma}\,\widetilde{\alpha}_{\bf k \sigma}$. 
In what follows, we shall fix the chemical potential to zero. 
The non-equilibrium evolution of the system is described by the conduction electron Green's function: 
$g_{\bf k \sigma}(t-t')=-i\langle {\cal T_C} \widetilde{\alpha}^{}_{\bf k \sigma}(t)
\widetilde{\alpha}^{\dagger}_{\bf k \sigma}(t')\rangle
$, 
where $\cal T_C$ is the time ordering operator on the Keldysh contour. Within this language,
we can define four Green's functions. Two combinations define the greater and lesser 
components
\begin{eqnarray} \label{Eq:g}
g^{>}_{\bf k \sigma}(t-t')&=&-i\langle \widetilde{\alpha}^{}_{\bf k \sigma}(t)\,
\widetilde{\alpha}^{\dagger}_{\bf k \sigma}(t')\rangle, \nonumber \\
g^{<}_{\bf k \sigma}(t-t')&=&i\langle \widetilde{\alpha}^{\dagger}_{\bf k \sigma}(t')\,
\widetilde{\alpha}^{}_{\bf k \sigma}(t)
\rangle,
\end{eqnarray}
while the other two define the time and anti-time ordered ones:
$
g^{t}_{\bf k \sigma}(t-t')= -i\langle {\cal T}\,\widetilde{\alpha}^{}_{\bf k \sigma}(t)\,
\widetilde{\alpha}^{\dagger}_{\bf k \sigma}(t')\rangle$, and $  
g^{\tilde t}_{\bf k \sigma}(t-t')=-i\langle\tilde {\cal  T}\widetilde{\alpha}^{}_{\bf k \sigma}(t)\,
\widetilde{\alpha}^{\dagger}_{\bf k \sigma}(t')\rangle$.  
Here ${\cal T}$ and $ \cal \tilde T$ are the time and anti-time ordering operators.
We also introduce the 
retarded and the advanced Green's functions, which are defined in the usual way:
\begin{eqnarray}
g^{A}_{\bf k \sigma}(t-t') & = & i \Theta (t'-t)\langle\{\widetilde{\alpha}^{}_{\bf k \sigma}(t),
\widetilde{\alpha}^{\dagger}_{\bf k \sigma}(t')\}\rangle,\nonumber \\
g^{R}_{\bf k \sigma}(t-t') & = & -i \Theta (t-t')\langle\{\widetilde{\alpha}^{}_{\bf k \sigma}(t),
\widetilde{\alpha}^{\dagger}_{\bf k \sigma}(t')\}\rangle.
\end{eqnarray}
We are interested in the momentum integrated Green's function
$g_{\sigma}(\omega)= \sum_{\bf k}g_{\bf k\s}(\omega)$, as these are the only quantities that 
explicitly enter 
the expression for the ac thermopower. Here, we consider the 
simplest situation of a dispersionless electronic band  
with a band cutoff D. It is characterized by a constant density of states 
$N(\omega) = 1/(2D)\, \Theta(D-|\omega|)= N(0)\, \Theta_{\omega}$, 
with $N(0)= 1/(2D)$, the DOS at the Fermi level 
and $\Theta_{\omega}=\Theta(D-|\omega|)$. 
Then, the momentum integrated 
Green's function have relatively simple analytical expressions~\cite{Koerting.07}:
$g^{>}_{\sigma}(\omega)=-2 \pi i\, (1-f(\omega))\,N(0)\,\Theta_{\omega}$,
$g^{<}_{\sigma}(\omega)=2 \pi i\, f(\omega)\,N(0)\,\Theta_{\omega}$ and
$\textrm{Im}\, g^{A}_{\sigma}(\omega)=-\textrm{Im}\, g^{R}_{\sigma}(\omega)  =  i\, \pi  
N(0)\Theta_{\omega} $.
Usually, $\textrm{Re}\, g^{R/A}$ is neglected in the large-bandwidth limit, but as
long as we are interested in the response functions $L_{ij}$ in the large $\omega>D$ limit, its contribution 
becomes important. 

\section{Current Correlations and the dynamical transport coefficients }\label{app:B}
\begin{figure}[tbh]
\includegraphics[width=0.49\columnwidth]{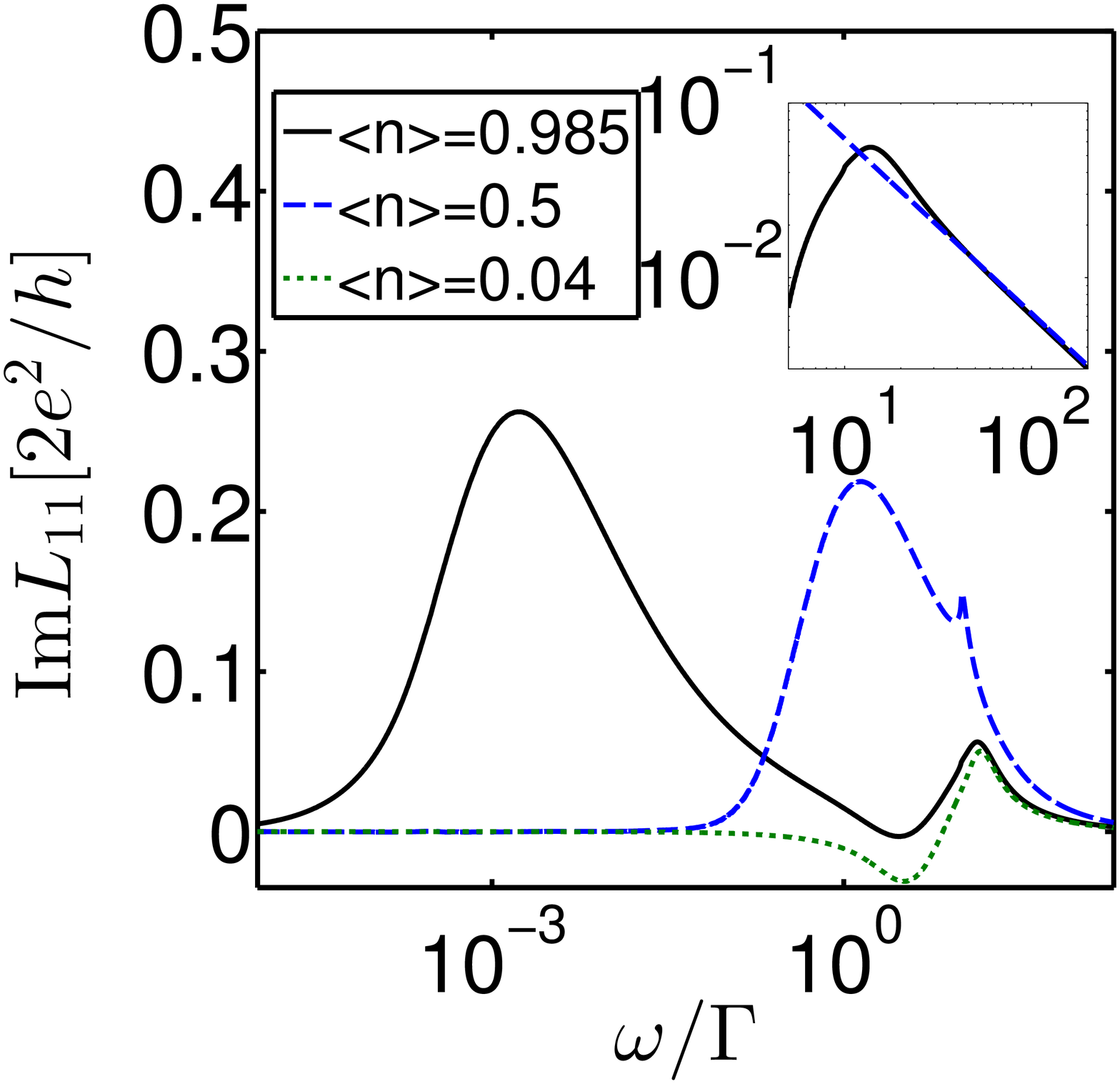}
\includegraphics[width=0.49\columnwidth]{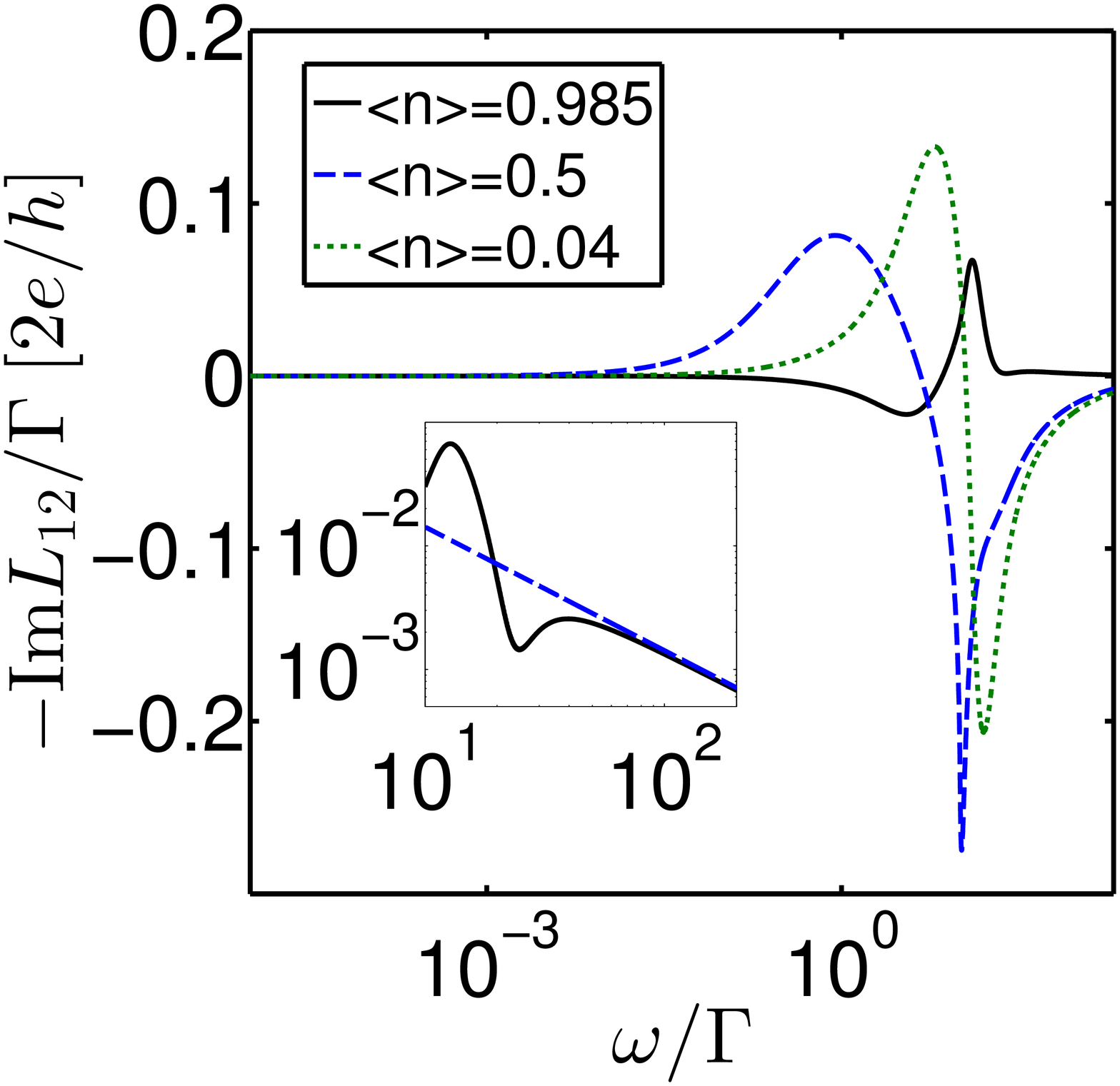}
\caption{(Color online) The imaginary part of the dynamical coefficients 
$L_{ij}(\omega)$ as function of frequency.  The inset shows the 
asymptotic $\sim 1/\omega$ behavior in the large-$\omega$ limit in the Kondo regime 
for $\langle n \rangle = 0.985$. The temperature was 
fixed to T=$10^{-5} \Gamma$.}
\label{fig:ImLij}
\end{figure}

\begin{figure}[h]
\includegraphics[clip,width=0.95\columnwidth]{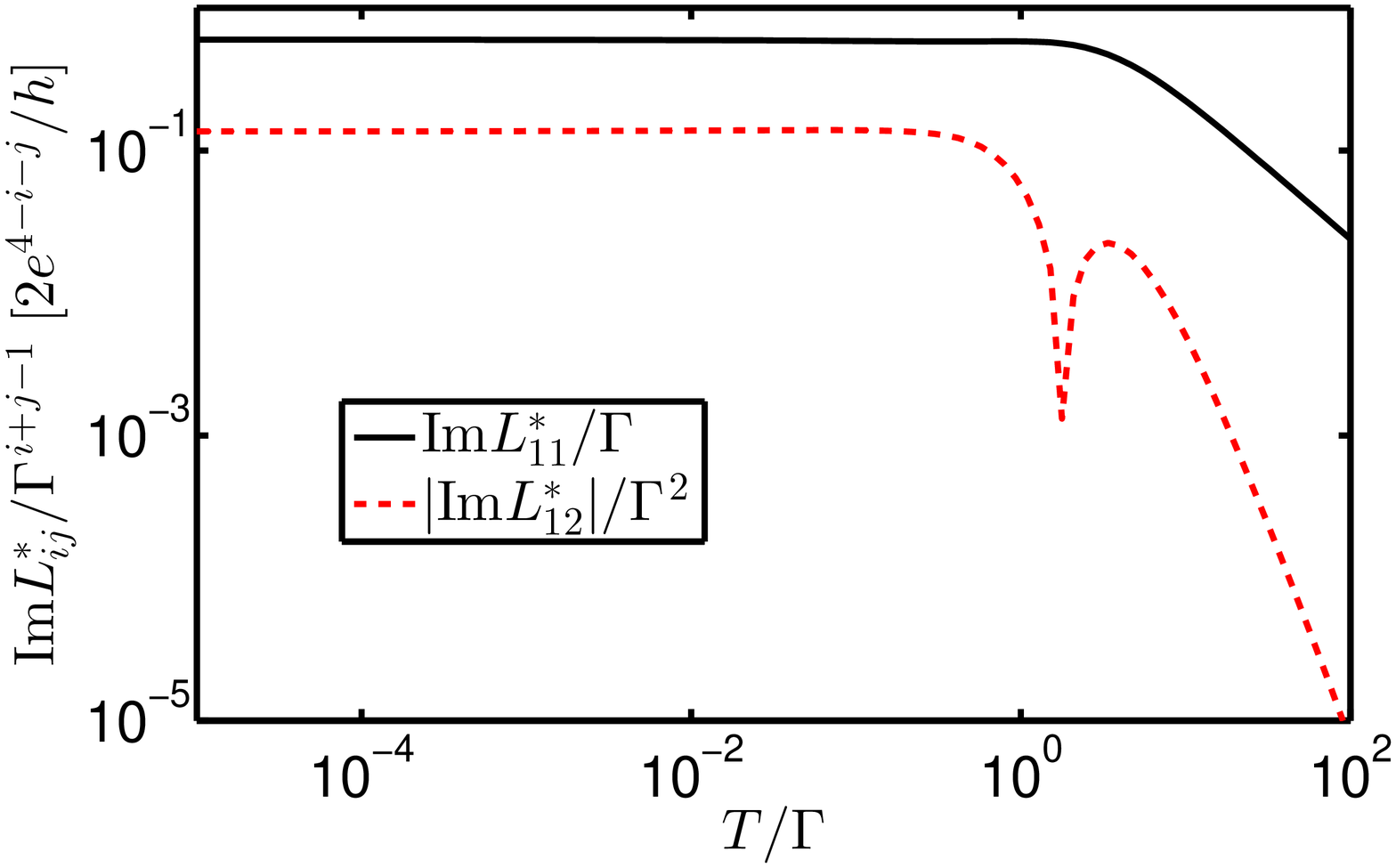}
\caption{(Color online) The temperature dependence of $L_{ij}^*$ in the Kondo regime. 
The results are obtained by using the sum rule expression, Eq.~\eqref{Eq:L_star}.}
\label{fig:L_star}
\end{figure}

\begin{figure}[t]
\includegraphics[width=0.49\columnwidth]{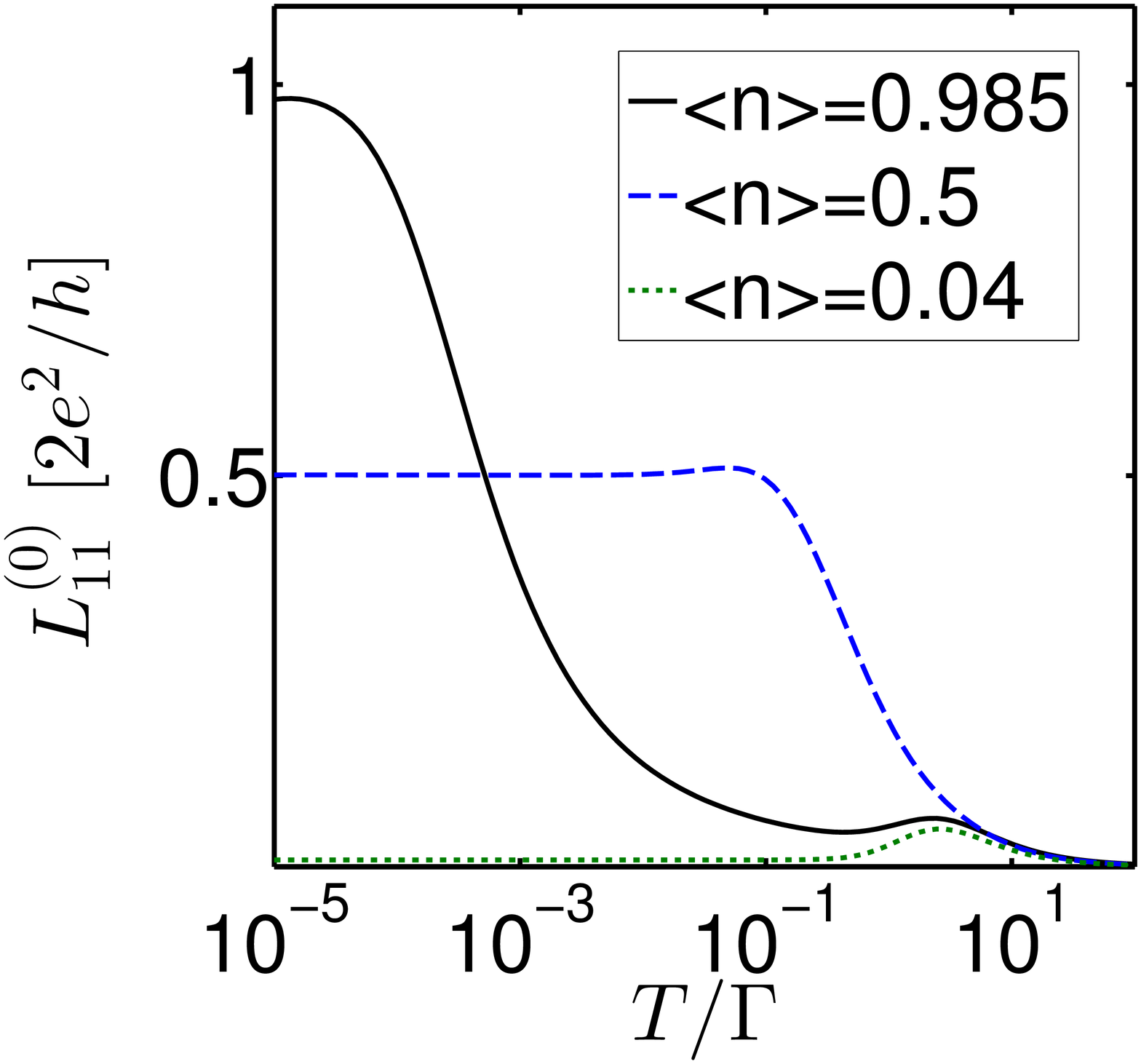}
\includegraphics[width=0.49\columnwidth]{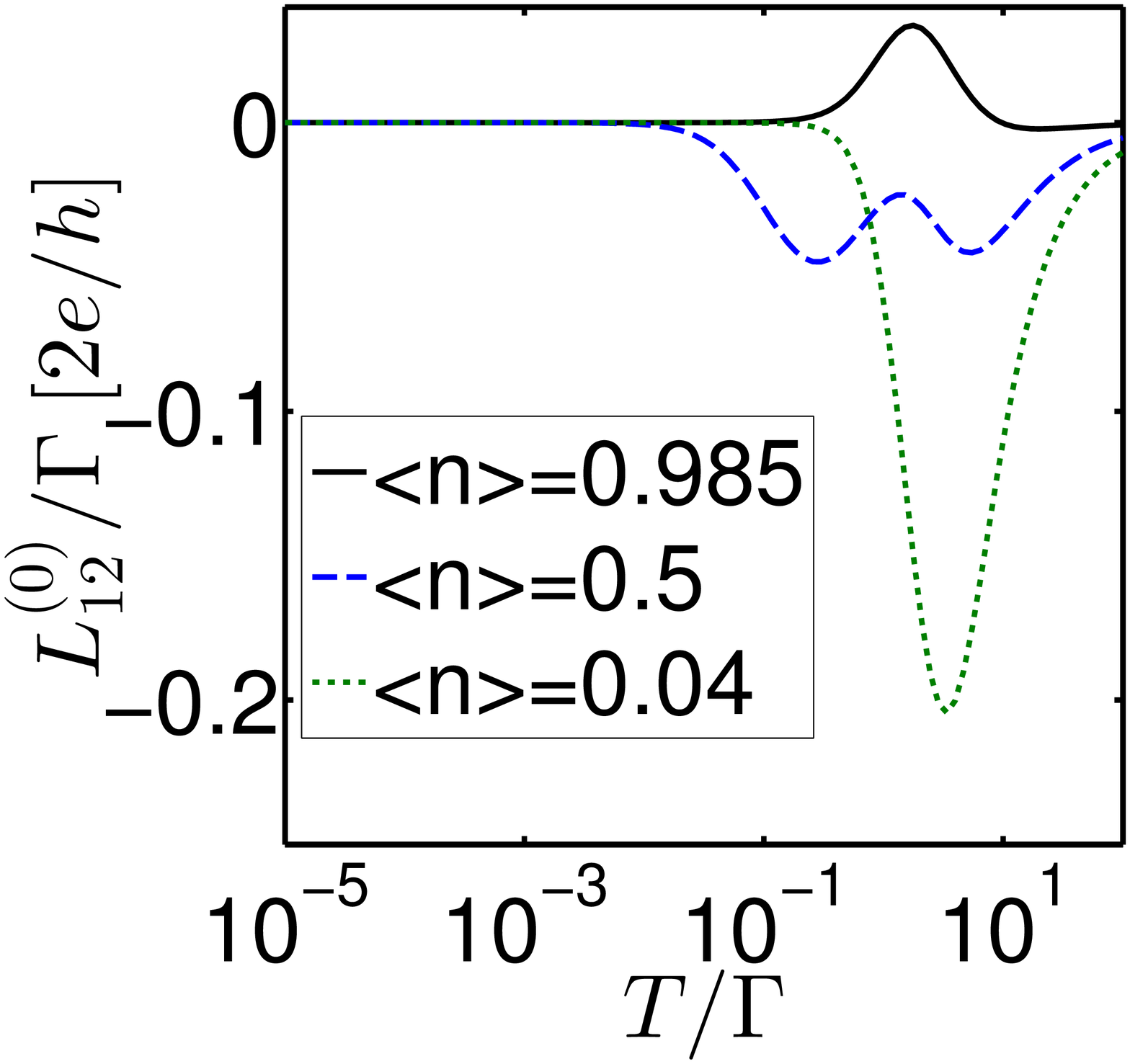}
\caption{(Color online). The temperature dependence of $\rm Re\, L_{ij}^{(0)}$. Here  
$L_{11}^{(0)}$ can be identified with the 
dc-conductance itself.}
\label{fig:ReL0}
\end{figure}

To get the dynamical transport coefficients, we need to compute the  
retarded response functions defined in 
Eq. \eqref{Eq:T}, with the current operators defined in Eqs.~\eqref{Eq:I_charge} and \eqref{Eq:I_heat}.
In the even/odd basis, one channel becomes decoupled and  
the charge current correlator $T_{11}$ can be evaluated as: 
\begin{gather}\label{Eq:T11}
T_{11}(t-t')=\Theta(t-t')\frac{e^2}{\hbar^3}\frac{t_{\rm eff}^2}{4}\sum_{\bf k, \sigma}  \nonumber \\
\left \{ G_{d}^{>}(t-t')g^{<}_{\bf k \sigma}(t'-t)-G_{d}^{<}(t-t')g^{>}_{\bf k \sigma}(t'-t)-\right.\nonumber \\
 \left. -G_d^{<}(t'-t)g^{>}_{\bf k \sigma}(t-t')-G_d^{>}(t'-t)g^{<}_{\bf k \sigma}(t-t') \right \}, 
 \end{gather}
which gives for  $L_{11}$  defined in Eq.~\eqref{Eq:L} a relatively compact expression:
\begin{gather} \label{Eq:ReL11}
\textrm{Re}\,L_{11}(\omega)=-\frac{t_{\rm eff}^2}{2\omega} \frac{e^2}{\hbar^3} \varrho_0
\int \rm d\omega'\left \{\textrm{Im}G_d^R(\omega')\left[\right. \right.\nonumber \\ 
\Theta_{\omega'-\omega}f(\omega'-\omega) + \Theta_{\omega'+\omega}f(\omega')-\Theta_{\omega'-\omega}f(\omega')-
\nonumber \\ 
\left. \left.-\Theta_{\omega'+\omega}f(\omega'+\omega)\right]  \right\}.
\end{gather}
Notice that within the present formalism, we can identify $\rm Re\, L_{11}(\omega)$
with the real part of the ac conductance $G(\omega)$.~\cite{Sindel.05}  
A similar expression 
can be derived for $T_{12}(t,t')$, which in the Fourier space becomes:
\begin{gather} \label{Eq:T12}
T_{12}(\omega)=-\frac{t_{\rm eff}^2}{4}\frac{e}{\hbar^3}\sum_{\bf k, \sigma}
\varepsilon_{\bf k}\int\frac{\rm d \omega'}{2\pi}\left\{G_d^R(\omega+\omega') g_{\bf k \sigma}^{<}(\omega')
\right. \nonumber \\
 +G_d^{<}(\omega+\omega')g_{\bf k \sigma}^{A}(\omega')
 +G_d^{>}(\omega')g_{\bf k \sigma}^{R}(\omega+\omega')+  \nonumber \\
\left.+G_d^{A}(\omega')g_{\bf k \sigma}^{>}(\omega+\omega')\right \}.
\end{gather}
Subtracting the $T_{12}(\omega=0)$ term and dividing by $-i\omega$, the real part of 
$L_{12}$ is obtained as:
\begin{gather} \label{Eq:ReL12}
\textrm{Re}\,L_{12}(\omega)=\frac{t_{\rm eff}^2}{2\omega} \frac{e}{\hbar^2}
\varrho_0
\int \rm d\omega'\left \{\textrm{Im}G_d^R(\omega')\left[\right. \right.\nonumber \\ 
(\omega'-\omega)\Theta_{\omega'-\omega}f(\omega'-\omega) + (\omega'+\omega)\Theta_{\omega'+\omega}f(\omega')-\nonumber \\
\left. \left.-(\omega'-\omega)\Theta_{\omega'-\omega}f(\omega')
-(\omega'+\omega)\Theta_{\omega'+\omega}f(\omega'+\omega)\right]  \right\}.
\end{gather}

In these expressions, $L_{ij}$ depends explicitly on the retarded localized d-level Green's function $\rm G_d^R(\omega)$. This quantity shall be computed exactly by using the NRG method.
In this way, Eqs. \eqref{Eq:ReL11} and \eqref{Eq:ReL12} are the exact expressions for the
real parts of the Onsager transport coefficients, and no approximation of any kind was 
used so far. 
The ac thermopower depends not only on the real, but also on the imaginary parts of $L_{ij}$. 
Actually their
imaginary parts give the main contribution in the large-frequency limit. To obtain 
them, we use the Kramers-Kr\"onig relations, Eq.~\eqref{Eq:KK}. 
In the ac limit, when $\omega$ is the largest energy scale ($\omega\gg D$), we can simplify 
considerably the calculation by noticing that
\begin{equation} \label{Eq:asymptotic}
\textrm{Im}\,L_{ij}(\omega)\simeq \frac{L_{ij}^{*}}{\omega} 
\end{equation}
 with $L_{ij}^*$ some coefficients, 
\begin{equation}\label{Eq:L_star}
L_{ij}^*=\frac{1}{\pi}\int_{-\infty}^{+\infty}\textrm{Re}\,L_{ij}(\omega')\rm d\omega' ,
\end{equation}
which are thus determined as the sum rule of dynamical quantities.~\cite{Xu.11} 
In Fig.~\ref{fig:ImLij} we represent the imaginary parts of $L_{ij}$ in the Kondo regime, 
as computed by doing the KK transformations of Eq.~\eqref{Eq:ReL11} and \eqref{Eq:ReL12}.
The insets display the large-frequency behavior, which indicates that our approximation,
Eq.~\eqref{Eq:asymptotic}, is indeed correct. The temperature dependence of $L_{ij}^*$ is displayed
in Fig.~\ref{fig:L_star}. 

In the small-frequency limit, 
$\omega\rightarrow 0$, the calculation can be simplified again. 
Introducing the notation 
$\textrm{Re}\, L_{ij}^{(0)}=\textrm{Re}L_{ij}\,(\omega\rightarrow 0)$,
we notice that Eqs.~\eqref{Eq:ReL11} and \eqref{Eq:ReL12} reduce considerably to
\begin{multline} \label{Eq:KK_L}
\textrm{Re}\,L_{ij}^{(0)} =   \frac{t_{\rm eff}^2}{\hbar}\left( -\frac{e}{\hbar}\right)^{4-i-j}  
\varrho_0\times\,\\ 
\int \rm d\tilde \omega\, \left ( \tilde \omega ^{i+j-2} \textrm{Im}G_d^R(\tilde \omega)
\frac{\partial f(\tilde \omega)}{\partial \tilde \omega}\right ).
\end{multline}
Here, $\textrm{Re}\, L_{11}^{(0)}$ is the dc-conductance itself. Its temperature dependence
is displayed in Fig.~\ref{fig:ReL0}. As expected in the $T\rightarrow 0$ limit, $\textrm{Re}\, 
L_{11}^{(0)}$ shows the usual Kondo behavior, as the system is close to the 
unitary limit. From Eqs. (\ref{Eq:KK_L}) and (\ref{Eq:S_omega}) one can notice that the thermopower has the form of an average entropy $-\left<\epsilon-\mu\right>/(eT)$ carried per particle/hole across the quantum dot. Thus, for the case of perfect particle-hole symmetry, the thermopower becomes zero. This form can also be used to justify the  $1/T$ dependence of $S^*$ (see Fig. \ref{fig:S}).





\bibliography{references}

\begin{thebibliography}{35}%
\makeatletter
\providecommand \@ifxundefined [1]{%
 \@ifx{#1\undefined}
}%
\providecommand \@ifnum [1]{%
 \ifnum #1\expandafter \@firstoftwo
 \else \expandafter \@secondoftwo
 \fi
}%
\providecommand \@ifx [1]{%
 \ifx #1\expandafter \@firstoftwo
 \else \expandafter \@secondoftwo
 \fi
}%
\providecommand \natexlab [1]{#1}%
\providecommand \enquote  [1]{``#1''}%
\providecommand \bibnamefont  [1]{#1}%
\providecommand \bibfnamefont [1]{#1}%
\providecommand \citenamefont [1]{#1}%
\providecommand \href@noop [0]{\@secondoftwo}%
\providecommand \href [0]{\begingroup \@sanitize@url \@href}%
\providecommand \@href[1]{\@@startlink{#1}\@@href}%
\providecommand \@@href[1]{\endgroup#1\@@endlink}%
\providecommand \@sanitize@url [0]{\catcode `\\12\catcode `\$12\catcode
  `\&12\catcode `\#12\catcode `\^12\catcode `\_12\catcode `\%12\relax}%
\providecommand \@@startlink[1]{}%
\providecommand \@@endlink[0]{}%
\providecommand \url  [0]{\begingroup\@sanitize@url \@url }%
\providecommand \@url [1]{\endgroup\@href {#1}{\urlprefix }}%
\providecommand \urlprefix  [0]{URL }%
\providecommand \Eprint [0]{\href }%
\providecommand \doibase [0]{http://dx.doi.org/}%
\providecommand \selectlanguage [0]{\@gobble}%
\providecommand \bibinfo  [0]{\@secondoftwo}%
\providecommand \bibfield  [0]{\@secondoftwo}%
\providecommand \translation [1]{[#1]}%
\providecommand \BibitemOpen [0]{}%
\providecommand \bibitemStop [0]{}%
\providecommand \bibitemNoStop [0]{.\EOS\space}%
\providecommand \EOS [0]{\spacefactor3000\relax}%
\providecommand \BibitemShut  [1]{\csname bibitem#1\endcsname}%
\let\auto@bib@innerbib\@empty
\bibitem [{\citenamefont {Sales}\ \emph {et~al.}(1996)\citenamefont {Sales},
  \citenamefont {Mandrus},\ and\ \citenamefont {Williams}}]{Sales.96}%
  \BibitemOpen
  \bibfield  {author} {\bibinfo {author} {\bibfnamefont {B.~C.}\ \bibnamefont
  {Sales}}, \bibinfo {author} {\bibfnamefont {D.}~\bibnamefont {Mandrus}}, \
  and\ \bibinfo {author} {\bibfnamefont {R.~K.}\ \bibnamefont {Williams}},\
  }\href@noop {} {\bibfield  {journal} {\bibinfo  {journal} {Science}\ }\textbf
  {\bibinfo {volume} {272}},\ \bibinfo {pages} {1325} (\bibinfo {year}
  {1996})}\BibitemShut {NoStop}%
\bibitem [{\citenamefont {Snyder}\ and\ \citenamefont
  {Toberer}(2008)}]{Snyder.08}%
  \BibitemOpen
  \bibfield  {author} {\bibinfo {author} {\bibfnamefont {G.~J.}\ \bibnamefont
  {Snyder}}\ and\ \bibinfo {author} {\bibfnamefont {E.~S.}\ \bibnamefont
  {Toberer}},\ }\href@noop {} {\bibfield  {journal} {\bibinfo  {journal} {Nat.
  Mater.}\ }\textbf {\bibinfo {volume} {7}},\ \bibinfo {pages} {104} (\bibinfo
  {year} {2008})}\BibitemShut {NoStop}%
\bibitem [{\citenamefont {Heremans}\ \emph {et~al.}(2008)\citenamefont
  {Heremans}, \citenamefont {Jovovic}, \citenamefont {Toberer}, \citenamefont
  {Saramat}, \citenamefont {Kurosaki}, \citenamefont {Charoenphakdee},
  \citenamefont {Yamanaka},\ and\ \citenamefont {Snyder}}]{Heremans.08}%
  \BibitemOpen
  \bibfield  {author} {\bibinfo {author} {\bibfnamefont {J.~P.}\ \bibnamefont
  {Heremans}}, \bibinfo {author} {\bibfnamefont {V.}~\bibnamefont {Jovovic}},
  \bibinfo {author} {\bibfnamefont {E.~S.}\ \bibnamefont {Toberer}}, \bibinfo
  {author} {\bibfnamefont {A.}~\bibnamefont {Saramat}}, \bibinfo {author}
  {\bibfnamefont {K.}~\bibnamefont {Kurosaki}}, \bibinfo {author}
  {\bibfnamefont {A.}~\bibnamefont {Charoenphakdee}}, \bibinfo {author}
  {\bibfnamefont {S.}~\bibnamefont {Yamanaka}}, \ and\ \bibinfo {author}
  {\bibfnamefont {G.~J.}\ \bibnamefont {Snyder}},\ }\href@noop {} {\bibfield
  {journal} {\bibinfo  {journal} {Science}\ }\textbf {\bibinfo {volume}
  {321}},\ \bibinfo {pages} {554} (\bibinfo {year} {2008})}\BibitemShut
  {NoStop}%
\bibitem [{\citenamefont {Humphrey}\ \emph {et~al.}(2002)\citenamefont
  {Humphrey}, \citenamefont {Newbury}, \citenamefont {Taylor},\ and\
  \citenamefont {Linke}}]{Humphrey.02}%
  \BibitemOpen
  \bibfield  {author} {\bibinfo {author} {\bibfnamefont {T.~E.}\ \bibnamefont
  {Humphrey}}, \bibinfo {author} {\bibfnamefont {R.}~\bibnamefont {Newbury}},
  \bibinfo {author} {\bibfnamefont {R.~P.}\ \bibnamefont {Taylor}}, \ and\
  \bibinfo {author} {\bibfnamefont {H.}~\bibnamefont {Linke}},\ }\href@noop {}
  {\bibfield  {journal} {\bibinfo  {journal} {Phys. Rev. Lett.}\ }\textbf
  {\bibinfo {volume} {89}},\ \bibinfo {pages} {116801} (\bibinfo {year}
  {2002})}\BibitemShut {NoStop}%
\bibitem [{\citenamefont {Harman}\ \emph {et~al.}(2002)\citenamefont {Harman},
  \citenamefont {Taylor}, \citenamefont {Walsh},\ and\ \citenamefont
  {LaForge}}]{Harman.02}%
  \BibitemOpen
  \bibfield  {author} {\bibinfo {author} {\bibfnamefont {T.~C.}\ \bibnamefont
  {Harman}}, \bibinfo {author} {\bibfnamefont {P.~J.}\ \bibnamefont {Taylor}},
  \bibinfo {author} {\bibfnamefont {M.~P.}\ \bibnamefont {Walsh}}, \ and\
  \bibinfo {author} {\bibfnamefont {B.~E.}\ \bibnamefont {LaForge}},\
  }\href@noop {} {\bibfield  {journal} {\bibinfo  {journal} {Science}\ }\textbf
  {\bibinfo {volume} {297}},\ \bibinfo {pages} {2229} (\bibinfo {year}
  {2002})}\BibitemShut {NoStop}%
\bibitem [{\citenamefont {Wang}\ \emph {et~al.}(2008)\citenamefont {Wang},
  \citenamefont {Feser}, \citenamefont {Lee}, \citenamefont {Talapin},
  \citenamefont {Segalman},\ and\ \citenamefont {Majumdar}}]{Wang.08}%
  \BibitemOpen
  \bibfield  {author} {\bibinfo {author} {\bibfnamefont {R.~Y.}\ \bibnamefont
  {Wang}}, \bibinfo {author} {\bibfnamefont {J.~P.}\ \bibnamefont {Feser}},
  \bibinfo {author} {\bibfnamefont {J.-S.}\ \bibnamefont {Lee}}, \bibinfo
  {author} {\bibfnamefont {D.~V.}\ \bibnamefont {Talapin}}, \bibinfo {author}
  {\bibfnamefont {R.}~\bibnamefont {Segalman}}, \ and\ \bibinfo {author}
  {\bibfnamefont {A.}~\bibnamefont {Majumdar}},\ }\href@noop {} {\bibfield
  {journal} {\bibinfo  {journal} {Nano Letters}\ }\textbf {\bibinfo {volume}
  {8}},\ \bibinfo {pages} {2283} (\bibinfo {year} {2008})}\BibitemShut
  {NoStop}%
\bibitem [{\citenamefont {Svensson}\ \emph {et~al.}(2012)\citenamefont
  {Svensson}, \citenamefont {Persson}, \citenamefont {Hoffmann}, \citenamefont
  {Nakpathomkun}, \citenamefont {Nilsson}, \citenamefont {Xu}, \citenamefont
  {Samuelson},\ and\ \citenamefont {Linke}}]{Svensson.12}%
  \BibitemOpen
  \bibfield  {author} {\bibinfo {author} {\bibfnamefont {S.~F.}\ \bibnamefont
  {Svensson}}, \bibinfo {author} {\bibfnamefont {A.~I.}\ \bibnamefont
  {Persson}}, \bibinfo {author} {\bibfnamefont {E.~A.}\ \bibnamefont
  {Hoffmann}}, \bibinfo {author} {\bibfnamefont {N.}~\bibnamefont
  {Nakpathomkun}}, \bibinfo {author} {\bibfnamefont {H.~A.}\ \bibnamefont
  {Nilsson}}, \bibinfo {author} {\bibfnamefont {H.~Q.}\ \bibnamefont {Xu}},
  \bibinfo {author} {\bibfnamefont {L.}~\bibnamefont {Samuelson}}, \ and\
  \bibinfo {author} {\bibfnamefont {H.}~\bibnamefont {Linke}},\ }\href@noop {}
  {\bibfield  {journal} {\bibinfo  {journal} {New Journal of Physics}\ }\textbf
  {\bibinfo {volume} {14}},\ \bibinfo {pages} {033041} (\bibinfo {year}
  {2012})}\BibitemShut {NoStop}%
\bibitem [{\citenamefont {Beenakker}\ and\ \citenamefont
  {Staring}(1992)}]{Beenakker.92}%
  \BibitemOpen
  \bibfield  {author} {\bibinfo {author} {\bibfnamefont {C.~W.~J.}\
  \bibnamefont {Beenakker}}\ and\ \bibinfo {author} {\bibfnamefont {A.~A.~M.}\
  \bibnamefont {Staring}},\ }\href@noop {} {\bibfield  {journal} {\bibinfo
  {journal} {Phys. Rev. B}\ }\textbf {\bibinfo {volume} {46}},\ \bibinfo
  {pages} {9667} (\bibinfo {year} {1992})}\BibitemShut {NoStop}%
\bibitem [{\citenamefont {Boese}\ and\ \citenamefont {Fazio}(2001)}]{Boese.01}%
  \BibitemOpen
  \bibfield  {author} {\bibinfo {author} {\bibfnamefont {D.}~\bibnamefont
  {Boese}}\ and\ \bibinfo {author} {\bibfnamefont {R.}~\bibnamefont {Fazio}},\
  }\href@noop {} {\bibfield  {journal} {\bibinfo  {journal} {Europhys. Lett.}\
  }\textbf {\bibinfo {volume} {56}},\ \bibinfo {pages} {576} (\bibinfo {year}
  {2001})}\BibitemShut {NoStop}%
\bibitem [{\citenamefont {Scheibner}\ \emph {et~al.}(2005)\citenamefont
  {Scheibner}, \citenamefont {Buhmann}, \citenamefont {Reuter}, \citenamefont
  {Kiselev},\ and\ \citenamefont {Molenkamp}}]{Scheibner.05}%
  \BibitemOpen
  \bibfield  {author} {\bibinfo {author} {\bibfnamefont {R.}~\bibnamefont
  {Scheibner}}, \bibinfo {author} {\bibfnamefont {H.}~\bibnamefont {Buhmann}},
  \bibinfo {author} {\bibfnamefont {D.}~\bibnamefont {Reuter}}, \bibinfo
  {author} {\bibfnamefont {M.~N.}\ \bibnamefont {Kiselev}}, \ and\ \bibinfo
  {author} {\bibfnamefont {L.~W.}\ \bibnamefont {Molenkamp}},\ }\href@noop {}
  {\bibfield  {journal} {\bibinfo  {journal} {Phys. Rev. Lett.}\ }\textbf
  {\bibinfo {volume} {95}},\ \bibinfo {pages} {176602} (\bibinfo {year}
  {2005})}\BibitemShut {NoStop}%
\bibitem [{\citenamefont {Costi}\ and\ \citenamefont
  {Zlati\ifmmode~\acute{c}\else \'{c}\fi{}}(2010)}]{Costi.10}%
  \BibitemOpen
  \bibfield  {author} {\bibinfo {author} {\bibfnamefont {T.~A.}\ \bibnamefont
  {Costi}}\ and\ \bibinfo {author} {\bibfnamefont {V.}~\bibnamefont
  {Zlati\ifmmode~\acute{c}\else \'{c}\fi{}}},\ }\href@noop {} {\bibfield
  {journal} {\bibinfo  {journal} {Phys. Rev. B}\ }\textbf {\bibinfo {volume}
  {81}},\ \bibinfo {pages} {235127} (\bibinfo {year} {2010})}\BibitemShut
  {NoStop}%
\bibitem [{\citenamefont {Roura-Bas}\ \emph {et~al.}(2012)\citenamefont
  {Roura-Bas}, \citenamefont {Tosi}, \citenamefont {Aligia},\ and\
  \citenamefont {Cornaglia}}]{Bas.12}%
  \BibitemOpen
  \bibfield  {author} {\bibinfo {author} {\bibfnamefont {P.}~\bibnamefont
  {Roura-Bas}}, \bibinfo {author} {\bibfnamefont {L.}~\bibnamefont {Tosi}},
  \bibinfo {author} {\bibfnamefont {A.~A.}\ \bibnamefont {Aligia}}, \ and\
  \bibinfo {author} {\bibfnamefont {P.~S.}\ \bibnamefont {Cornaglia}},\
  }\href@noop {} {\bibfield  {journal} {\bibinfo  {journal} {Phys. Rev. B}\
  }\textbf {\bibinfo {volume} {86}},\ \bibinfo {pages} {165106} (\bibinfo
  {year} {2012})}\BibitemShut {NoStop}%
\bibitem [{\citenamefont {Trocha}\ and\ \citenamefont
  {Barna\ifmmode~\acute{s}\else \'{s}\fi{}}(2012)}]{Trocha.12}%
  \BibitemOpen
  \bibfield  {author} {\bibinfo {author} {\bibfnamefont {P.}~\bibnamefont
  {Trocha}}\ and\ \bibinfo {author} {\bibfnamefont {J.}~\bibnamefont
  {Barna\ifmmode~\acute{s}\else \'{s}\fi{}}},\ }\href@noop {} {\bibfield
  {journal} {\bibinfo  {journal} {Phys. Rev. B}\ }\textbf {\bibinfo {volume}
  {85}},\ \bibinfo {pages} {085408} (\bibinfo {year} {2012})}\BibitemShut
  {NoStop}%
\bibitem [{\citenamefont {Donsa}\ \emph {et~al.}(2013)\citenamefont {Donsa},
  \citenamefont {Andergassen},\ and\ \citenamefont {Held}}]{Donsa.13}%
  \BibitemOpen
  \bibfield  {author} {\bibinfo {author} {\bibfnamefont {S.}~\bibnamefont
  {Donsa}}, \bibinfo {author} {\bibfnamefont {S.}~\bibnamefont {Andergassen}},
  \ and\ \bibinfo {author} {\bibfnamefont {K.}~\bibnamefont {Held}},\
  }\href@noop {} {\bibfield  {journal} {\bibinfo  {journal} {arXiv:1308.4882}\
  } (\bibinfo {year} {2013})}\BibitemShut {NoStop}%
\bibitem [{\citenamefont {Lim}\ \emph {et~al.}(2013)\citenamefont {Lim},
  \citenamefont {L\'opez},\ and\ \citenamefont {S\'anchez}}]{Lopez.13}%
  \BibitemOpen
  \bibfield  {author} {\bibinfo {author} {\bibfnamefont {J.~S.}\ \bibnamefont
  {Lim}}, \bibinfo {author} {\bibfnamefont {R.}~\bibnamefont {L\'opez}}, \ and\
  \bibinfo {author} {\bibfnamefont {D.}~\bibnamefont {S\'anchez}},\ }\href@noop
  {} {\bibfield  {journal} {\bibinfo  {journal} {Phys. Rev. B}\ }\textbf
  {\bibinfo {volume} {88}},\ \bibinfo {pages} {201304} (\bibinfo {year}
  {2013})}\BibitemShut {NoStop}%
\bibitem [{\citenamefont {Sindel}\ \emph {et~al.}(2005)\citenamefont {Sindel},
  \citenamefont {Hofstetter}, \citenamefont {von Delft},\ and\ \citenamefont
  {Kindermann}}]{Sindel.05}%
  \BibitemOpen
  \bibfield  {author} {\bibinfo {author} {\bibfnamefont {M.}~\bibnamefont
  {Sindel}}, \bibinfo {author} {\bibfnamefont {W.}~\bibnamefont {Hofstetter}},
  \bibinfo {author} {\bibfnamefont {J.}~\bibnamefont {von Delft}}, \ and\
  \bibinfo {author} {\bibfnamefont {M.}~\bibnamefont {Kindermann}},\
  }\href@noop {} {\bibfield  {journal} {\bibinfo  {journal} {Phys. Rev. Lett.}\
  }\textbf {\bibinfo {volume} {94}},\ \bibinfo {pages} {196602} (\bibinfo
  {year} {2005})}\BibitemShut {NoStop}%
\bibitem [{\citenamefont {Blanter}\ and\ \citenamefont
  {Sukhorukov}(2000)}]{Blanter.00}%
  \BibitemOpen
  \bibfield  {author} {\bibinfo {author} {\bibfnamefont {Y.~M.}\ \bibnamefont
  {Blanter}}\ and\ \bibinfo {author} {\bibfnamefont {E.~V.}\ \bibnamefont
  {Sukhorukov}},\ }\href@noop {} {\bibfield  {journal} {\bibinfo  {journal}
  {Phys. Rev. Lett.}\ }\textbf {\bibinfo {volume} {84}},\ \bibinfo {pages}
  {1280} (\bibinfo {year} {2000})}\BibitemShut {NoStop}%
\bibitem [{\citenamefont {Moca}\ \emph {et~al.}(2011)\citenamefont {Moca},
  \citenamefont {Weymann},\ and\ \citenamefont {Zarand}}]{Moca.11}%
  \BibitemOpen
  \bibfield  {author} {\bibinfo {author} {\bibfnamefont {C.~P.}\ \bibnamefont
  {Moca}}, \bibinfo {author} {\bibfnamefont {I.}~\bibnamefont {Weymann}}, \
  and\ \bibinfo {author} {\bibfnamefont {G.}~\bibnamefont {Zarand}},\
  }\href@noop {} {\bibfield  {journal} {\bibinfo  {journal} {Phys. Rev. B}\
  }\textbf {\bibinfo {volume} {84}},\ \bibinfo {pages} {235441} (\bibinfo
  {year} {2011})}\BibitemShut {NoStop}%
\bibitem [{\citenamefont {Kogan}\ \emph {et~al.}(2004)\citenamefont {Kogan},
  \citenamefont {Amasha},\ and\ \citenamefont {Kastner}}]{Kogan.04}%
  \BibitemOpen
  \bibfield  {author} {\bibinfo {author} {\bibfnamefont {A.}~\bibnamefont
  {Kogan}}, \bibinfo {author} {\bibfnamefont {S.}~\bibnamefont {Amasha}}, \
  and\ \bibinfo {author} {\bibfnamefont {M.~A.}\ \bibnamefont {Kastner}},\
  }\href@noop {} {\bibfield  {journal} {\bibinfo  {journal} {Science}\ }\textbf
  {\bibinfo {volume} {304}},\ \bibinfo {pages} {1293} (\bibinfo {year}
  {2004})}\BibitemShut {NoStop}%
\bibitem [{\citenamefont {Basset}\ \emph {et~al.}(2012)\citenamefont {Basset},
  \citenamefont {Kasumov}, \citenamefont {Moca}, \citenamefont {Zar\'and},
  \citenamefont {Simon}, \citenamefont {Bouchiat},\ and\ \citenamefont
  {Deblock}}]{Basset.12}%
  \BibitemOpen
  \bibfield  {author} {\bibinfo {author} {\bibfnamefont {J.}~\bibnamefont
  {Basset}}, \bibinfo {author} {\bibfnamefont {A.~Y.}\ \bibnamefont {Kasumov}},
  \bibinfo {author} {\bibfnamefont {C.~P.}\ \bibnamefont {Moca}}, \bibinfo
  {author} {\bibfnamefont {G.}~\bibnamefont {Zar\'and}}, \bibinfo {author}
  {\bibfnamefont {P.}~\bibnamefont {Simon}}, \bibinfo {author} {\bibfnamefont
  {H.}~\bibnamefont {Bouchiat}}, \ and\ \bibinfo {author} {\bibfnamefont
  {R.}~\bibnamefont {Deblock}},\ }\href@noop {} {\bibfield  {journal} {\bibinfo
   {journal} {Phys. Rev. Lett.}\ }\textbf {\bibinfo {volume} {108}},\ \bibinfo
  {pages} {046802} (\bibinfo {year} {2012})}\BibitemShut {NoStop}%
\bibitem [{\citenamefont {Shastry}(2006)}]{Shastry.06}%
  \BibitemOpen
  \bibfield  {author} {\bibinfo {author} {\bibfnamefont {B.~S.}\ \bibnamefont
  {Shastry}},\ }\href@noop {} {\bibfield  {journal} {\bibinfo  {journal} {Phys.
  Rev. B}\ }\textbf {\bibinfo {volume} {73}},\ \bibinfo {pages} {085117}
  (\bibinfo {year} {2006})}\BibitemShut {NoStop}%
\bibitem [{\citenamefont {Shastry}(2009)}]{Shastry.09}%
  \BibitemOpen
  \bibfield  {author} {\bibinfo {author} {\bibfnamefont {B.~S.}\ \bibnamefont
  {Shastry}},\ }\href@noop {} {\bibfield  {journal} {\bibinfo  {journal}
  {Reports on Progress in Physics}\ }\textbf {\bibinfo {volume} {72}},\
  \bibinfo {pages} {016501} (\bibinfo {year} {2009})}\BibitemShut {NoStop}%
\bibitem [{\citenamefont {Xu}\ \emph {et~al.}(2011)\citenamefont {Xu},
  \citenamefont {Weber},\ and\ \citenamefont {Kotliar}}]{Xu.11}%
  \BibitemOpen
  \bibfield  {author} {\bibinfo {author} {\bibfnamefont {W.}~\bibnamefont
  {Xu}}, \bibinfo {author} {\bibfnamefont {C.}~\bibnamefont {Weber}}, \ and\
  \bibinfo {author} {\bibfnamefont {G.}~\bibnamefont {Kotliar}},\ }\href@noop
  {} {\bibfield  {journal} {\bibinfo  {journal} {Phys. Rev. B}\ }\textbf
  {\bibinfo {volume} {84}},\ \bibinfo {pages} {035114} (\bibinfo {year}
  {2011})}\BibitemShut {NoStop}%
\bibitem [{Mah()}]{Mahan}%
  \BibitemOpen
  \href@noop {} {\bibinfo  {journal} {Mahan, G. D., {\it Many-Particle Physics}
  (Kluwer Academic, New York, 2000)}\ }\BibitemShut {NoStop}%
\bibitem [{\citenamefont {Luttinger}(1964)}]{Luttinger.64}%
  \BibitemOpen
\bibfield  {journal} {  }\bibfield  {author} {\bibinfo {author} {\bibfnamefont
  {J.~M.}\ \bibnamefont {Luttinger}},\ }\href@noop {} {\bibfield  {journal}
  {\bibinfo  {journal} {Phys. Rev.}\ }\textbf {\bibinfo {volume} {135}},\
  \bibinfo {pages} {A1505} (\bibinfo {year} {1964})}\BibitemShut {NoStop}%
\bibitem [{\citenamefont {Haldane}(1978)}]{Haldane.81}%
  \BibitemOpen
  \bibfield  {author} {\bibinfo {author} {\bibfnamefont {F.~D.~M.}\
  \bibnamefont {Haldane}},\ }\href@noop {} {\bibfield  {journal} {\bibinfo
  {journal} {Phys. Rev. Lett.}\ }\textbf {\bibinfo {volume} {40}},\ \bibinfo
  {pages} {416} (\bibinfo {year} {1978})}\BibitemShut {NoStop}%
\bibitem [{hew()}]{hewson}%
  \BibitemOpen
  \href@noop {} {\bibinfo  {journal} {A. C. Hewson, {\it The Kondo Problem to
  Heavy Fermions} (Cambridge University Press, Cambridge, 1993)}\ }\BibitemShut
  {NoStop}%
\bibitem [{\citenamefont {Krishna-murthy}\ \emph
  {et~al.}(1980{\natexlab{a}})\citenamefont {Krishna-murthy}, \citenamefont
  {Wilkins},\ and\ \citenamefont {Wilson}}]{Wilson.80.1}%
  \BibitemOpen
\bibfield  {journal} {  }\bibfield  {author} {\bibinfo {author} {\bibfnamefont
  {H.~R.}\ \bibnamefont {Krishna-murthy}}, \bibinfo {author} {\bibfnamefont
  {J.~W.}\ \bibnamefont {Wilkins}}, \ and\ \bibinfo {author} {\bibfnamefont
  {K.~G.}\ \bibnamefont {Wilson}},\ }\href@noop {} {\bibfield  {journal}
  {\bibinfo  {journal} {Phys. Rev. B}\ }\textbf {\bibinfo {volume} {21}},\
  \bibinfo {pages} {1003} (\bibinfo {year} {1980}{\natexlab{a}})}\BibitemShut
  {NoStop}%
\bibitem [{\citenamefont {Krishna-murthy}\ \emph
  {et~al.}(1980{\natexlab{b}})\citenamefont {Krishna-murthy}, \citenamefont
  {Wilkins},\ and\ \citenamefont {Wilson}}]{Wilson.80.2}%
  \BibitemOpen
  \bibfield  {author} {\bibinfo {author} {\bibfnamefont {H.~R.}\ \bibnamefont
  {Krishna-murthy}}, \bibinfo {author} {\bibfnamefont {J.~W.}\ \bibnamefont
  {Wilkins}}, \ and\ \bibinfo {author} {\bibfnamefont {K.~G.}\ \bibnamefont
  {Wilson}},\ }\href@noop {} {\bibfield  {journal} {\bibinfo  {journal} {Phys.
  Rev. B}\ }\textbf {\bibinfo {volume} {21}},\ \bibinfo {pages} {1044}
  (\bibinfo {year} {1980}{\natexlab{b}})}\BibitemShut {NoStop}%
\bibitem [{\citenamefont {Bulla}\ \emph {et~al.}(2008)\citenamefont {Bulla},
  \citenamefont {Costi},\ and\ \citenamefont {Pruschke}}]{Bulla.08}%
  \BibitemOpen
  \bibfield  {author} {\bibinfo {author} {\bibfnamefont {R.}~\bibnamefont
  {Bulla}}, \bibinfo {author} {\bibfnamefont {T.~A.}\ \bibnamefont {Costi}}, \
  and\ \bibinfo {author} {\bibfnamefont {T.}~\bibnamefont {Pruschke}},\
  }\href@noop {} {\bibfield  {journal} {\bibinfo  {journal} {Rev. Mod. Phys.}\
  }\textbf {\bibinfo {volume} {80}},\ \bibinfo {pages} {395} (\bibinfo {year}
  {2008})}\BibitemShut {NoStop}%
\bibitem [{Bud()}]{BudapestNRG}%
  \BibitemOpen
  \href@noop {} {\bibinfo  {journal} {We used the open-access Budapest Flexible
  DM-NRG code, http://www.phy.bme.hu/\~{}dmnrg/; O. Legeza, C. P. Moca, A. I.
  T\'{o}th, I. Weymann, G. Zar\'{a}nd, arXiv:0809.3143 (2008) (unpublished)}\
  }\BibitemShut {NoStop}%
\bibitem [{\citenamefont {Nozières}(1974)}]{Nozieres.74}%
  \BibitemOpen
\bibfield  {journal} {  }\bibfield  {author} {\bibinfo {author} {\bibfnamefont
  {P.}~\bibnamefont {Nozières}},\ }\href@noop {} {\bibfield  {journal}
  {\bibinfo  {journal} {Journal of Low Temperature Physics}\ }\textbf {\bibinfo
  {volume} {17}},\ \bibinfo {pages} {31} (\bibinfo {year} {1974})}\BibitemShut
  {NoStop}%
\bibitem [{\citenamefont {Costi}\ and\ \citenamefont
  {Hewson}(1993)}]{Costi.93}%
  \BibitemOpen
  \bibfield  {author} {\bibinfo {author} {\bibfnamefont {T.~A.}\ \bibnamefont
  {Costi}}\ and\ \bibinfo {author} {\bibfnamefont {A.~C.}\ \bibnamefont
  {Hewson}},\ }\href@noop {} {\bibfield  {journal} {\bibinfo  {journal}
  {Journal of Physics: Condensed Matter}\ }\textbf {\bibinfo {volume} {5}},\
  \bibinfo {pages} {L361} (\bibinfo {year} {1993})}\BibitemShut {NoStop}%
\bibitem [{\citenamefont {Glazman}\ and\ \citenamefont
  {Pustilnik}(2003)}]{Glazman.03}%
  \BibitemOpen
  \bibfield  {author} {\bibinfo {author} {\bibfnamefont {L.}~\bibnamefont
  {Glazman}}\ and\ \bibinfo {author} {\bibfnamefont {M.}~\bibnamefont
  {Pustilnik}},\ }\bibfield  {booktitle} {\emph {\bibinfo {booktitle} {New
  Directions in Mesoscopic Physics (Towards Nanoscience)}},\ }\href@noop {} {\
  \bibinfo {series} {NATO Science Series},\ \textbf {\bibinfo {volume} {125}},\
  \bibinfo {pages} {93} (\bibinfo {year} {2003})}\BibitemShut {NoStop}%
\bibitem [{\citenamefont {K\"orting}()}]{Koerting.07}%
  \BibitemOpen
  \bibfield  {author} {\bibinfo {author} {\bibfnamefont {V.}~\bibnamefont
  {K\"orting}},\ }\href@noop {} {\bibinfo  {journal} {Ph.D. thesis, Karlsruhe
  University, 2007}\ }\BibitemShut {NoStop}%
\end{thebibliography}%

\end{document}